\documentclass[12pt]{article}

\usepackage{fullpage,times}

\long\def\comment#1{}

\def\tight{\itemsep=0pt\parsep=0pt\parskip=0pt}

\usepackage{natbib}           
\usepackage{graphicx}	
\usepackage{caption}
\usepackage{subcaption}
\usepackage{amsmath}
\usepackage{hyperref}
\usepackage{tikz}
\usetikzlibrary{fit,positioning,shapes,arrows, decorations.pathreplacing}
\usepackage{booktabs} 

\def\mathbb#1{\mbox{\boldmath$ #1$}}

\renewenvironment{cases}{\left\{\begin{array}{ll}}{\end{array}\right.}

\def\url#1{{\tt #1}}

\def\text#1{\mbox{#1}}

\def\logit{\mbox{logit}\,}

\definecolor{vpurple}{HTML}{440154}
\definecolor{vblue}{HTML}{3B528B}
\definecolor{vteal}{HTML}{21908C}
\definecolor{vgreen}{HTML}{5DC863}
\definecolor{vyellow}{HTML}{FDE725}

\defcitealias{nist2012latent}{NIST 2012}

\begin{document}

\title{Psychometric Analysis of Forensic Examiner Behavior\thanks{The material presented here is based upon work supported in part under Award No. 70NANB15H176 from the U.S. Department of Commerce, National Institute of Science and Technology. Any opinions, findings, or recommendations expressed in this material are those of the author(s) and do not necessarily reflect the views of the National Institute of Science and Technology, nor the Center for Statistics and Applications in Forensic Evidence.}}
\author{Amanda Luby\thanks{Swarthmore College}\\ 
\texttt{aluby1@swarthmore.edu}
\and Anjali Mazumder\thanks{The Alan Turing Institute, London}\\ 
\texttt{amazumder@turing.ac.uk}
\and Brian Junker\thanks{Carnegie Mellon University}\\
\texttt{brian@stat.cmu.edu}
}

\maketitle

\begin{abstract}
Forensic science often involves the comparison of crime-scene evidence to a known-source sample to determine if the evidence and the reference sample came from the same source. Even as forensic analysis tools become increasingly objective and automated, final source identifications are often left to individual examiners’ interpretation of the evidence.  Each source identification relies on judgements about the features and quality of the crime-scene evidence that may vary from one examiner to the next.  The current approach to characterizing uncertainty in examiners’ decision-making has largely centered around the calculation of error rates aggregated across examiners and identification tasks, without taking into account these variations in behavior. We propose a new approach using IRT and IRT-like models to account for differences among examiners and additionally account for the varying difficulty among source identification tasks. In particular, we survey some recent advances \citep{luby2019thesis} in the application of Bayesian psychometric models, including simple Rasch models as well as more elaborate decision tree models, to fingerprint examiner behavior.

\end{abstract}

\clearpage

\tableofcontents

\clearpage

\section{Introduction}\label{s:intro}

Validity and reliability of the evaluation of forensic science evidence is powerful and crucial to the fact-finding mission of the courts and criminal justice system \citep{pcast}. Common types of evidence include DNA taken from blood or tissue samples, glass fragments, shoe impressions, firearm bullets or casings, fingerprints, handwriting, and traces of online/digital behavior.  
Evaluating these types of evidence often involves comparing a crime scene sample, referred to in this field as a {\em latent} sample\footnote{This usage should not be confused with the usage of ``latent'' in psychometrics, meaning a variable related to individual differences that is unobservable.  We will use the word in both senses in this paper, the meaning being clear from context.}, with a sample from one or more persons of interest, referred to as {\em reference} samples; forensic scientists refer to this as an {\em identification task}.  Ideally, the result of an identification task is what is referred to as an {\em individualization}, i.e. an assessment by the examiner that the latent and reference samples come from the same source, or an {\em exclusion}, i.e. an assessment that the sources for the two samples are different.  For a variety of reasons, the assessments in identification tasks for some kinds of evidence can be much more accurate and precise than for others.

The evaluation and interpretation of forensic evidence often involve at least two steps: (a) comparing a latent sample to a reference sample, and (b) assessing the meaning of that reported match or non-match \citep{saks2008individualization}. There are often additional steps taken, for example, to assess whether the latent sample is of sufficient quality for comparison. Many kinds of identification tasks, e.g. those involving fingerprint, firearms and handwriting data, require human examiners to subjectively select features to compare in the latent and reference samples. The response provided by a forensic examiner is thus more nuanced than a dichotomous match or no-match decision.
Further, each of these steps introduces potential for variability and uncertainty by the forensic science examiner.  Finally, the latent samples can be of varying quality, contributing further to variability and uncertainty in completing identification tasks.  Forensic examination is thus ripe for the application of item response theory (IRT) and related psychometric models, in which examiners play the role of respondents or participants, and identification tasks play the role of items \citep{kerkhoff2015,luby2018proficiency}.

In this paper we survey recent advances in the psychometric analysis of forensic examiner behavior \citep{luby2019thesis}. In particular we will apply IRT 
and related models, including Rasch models \citep{rasch1960studies,raschbook}, models for collateral or covarying responses \citep[similar to][]{thissen_9_1983}, item response trees  \citep[IRTRees,][]{deboeck2012statsoft} and cultural consensus theory models \citep[CCT,][]{batchelder1988test}, to better understand the operating characteristics of identification tasks performed by human forensic examiners.  We will focus on fingerprint analysis, but the same techniques can be used to understand identification tasks for other types of forensic evidence.  Understanding examiners' performance is obviously of interest to legal decision makers, for whom the frequency and types of errors in forensic testimony is important \citep{garrett2017proficiency, max2019assessing}, but it can also lead to better pre-service and in-service training for examiners, to reduce erroneous or misleading testimony. 

\subsection{Fingerprint analysis}\label{ss:fingerprint-intro}

Fingerprint identification tasks in which an examiner compares a latent print to one or more reference prints involve many sources of variation and uncertainty.  The latent print may be smudged or otherwise degraded to varying degrees, making comparison with the reference print difficult or impossible.  The areas of the print available in the latent image may be difficult to locate in the reference print of interest.  Even the latent print is clear and complete, the degree of similarity between the latent and reference prints varies considerably across identification tasks. See, e.g. \citet{becue2019fingermarks} for a comprehensive review of fingerprint comparison.

Examiners also contribute variability and uncertainty to the process.   Different examiners may be differentially inclined in their determinations of whether print quality is sufficient to make a comparison.  They may choose different features, or {\em minutiae}, on which to base a comparison, and they may have different personal thresholds for similarity of individual minutiae, or for the number of minutiae that must match (respectively fail to match) to declare an individualization (respectively exclusion); see for example \citet{ulery2014measuring}.  

\subsection{Empirical work to date}\label{ss:previous-work}

Proficiency tests do exist for examiners \citep{pcast}, but they are typically scored with number-right or percent-correct scoring \citep{gardner2019latent}.  This approach does not account for differing difficulty of identification tasks across different editions of the same proficiency test, nor across tasks within a single proficiency test.  Thus the same score may indicate very different levels of examiner proficiency, depending on the difficulty of the tasks on a particular edition of the test, or even on the difficulty of the particular items answered correctly and incorrectly by different examiners with the same number-correct score on the same edition of the test. 

Error rate studies, that aggregate true-positive, true-negative, false-positive and false-negative rates across many examiners and identification tasks, contain unmeasured biases due to the above variations in task difficulty and examiner practice and proficiency; see for example \citet{luby2018proficiency}.  In addition, raw sample sizes in these studies understate true standard errors, due to correlation between responses from the same examiner \citep{holland1986conditional}.

\subsection{Preview}\label{ss:preview}

In this paper we review some recent advances \citep{luby2019thesis} in the application of Bayesian IRT and IRT-like models to fingerprint examiner proficiency testing and error rate data.  We show the additional information that can be obtained from application of even a simple IRT model \citep[e.g.,][]{rasch1960studies,raschbook} to proficiency data, and compare that information with examiners' perceived difficulty of identification tasks.  We also explore models for staged decision making and polytomous responses when there is no ground truth (answer key).  In this latter situation, even though there is no answer key, we are able to extract useful diagnostic information about examiners' decision processes, relative to a widely recommended decision process \citepalias[known as ACE-V,][]{nist2012latent}, using the IRTrees framework of \citet{deboeck2012statsoft}.  Interestingly the latent traits or person parameters in these models no longer represent proficiencies in performing identification tasks but rather tendencies of examiners toward one decision or another.   This leads to a better understanding of variation among examiners at different points in the analysis process.  Finally we compare the characteristics of IRT-like models for generating answer keys with the characteristics of social consensus models \citep{batchelder1988test, anders2015cultural} applied to the same problem.

\section{Available Forensic Data}\label{s:data}

The vast majority of forensic decision-making occurs in casework, which
is not often made available to researchers due to privacy concerns or
active investigation policies. Besides real-world casework, data on
forensic decision-making is collected through proficiency testing 
and error rate studies.  Proficiency tests are periodic
competency exams that must be completed for forensic laboratories to
maintain their accreditation, while error rate studies are research studies designed to measure casework error rates. 

\subsection{Proficiency Tests}\label{s:proficiency-exams}

Proficiency tests usually involve a large number of participants (often
\(>400\)), across multiple laboratories, responding to a small set of identification task items (often \(<20\)). Since
every participant responds to every item, we can assess participant
proficiency and item difficulty largely using the observed scores. Since
proficiency exams are designed to assess basic competency, most
items are relatively easy and the vast majority of participants
score 100\% on each test. 

In the US, forensic proficiency testing companies include {Collaborative Testing Services} (CTS), {Ron Smith and Associates} (RSA), {Forensic Testing Services} (FTS), and {Forensic Assurance} (FA). Both CTS and RSA provide two tests per year in fingerprint examination, consisting of 10-12 items, and make reports of the results available. FA also provides two tests per year, but does not provide reports of results. FTS does not offer proficiency tests for fingerprint examiners but instead focuses on other forensic domains.

\begin{figure}[h]
	\begin{subfigure}{.35\textwidth}
		 \centering
		 \includegraphics[width=.9\linewidth]{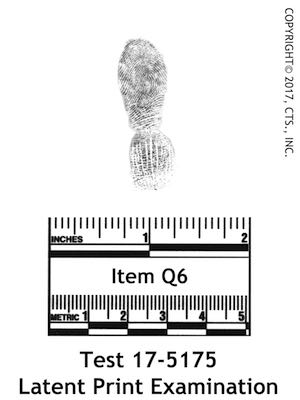}
		 \caption{A latent fingerprint sample provided by CTS.}
		\label{fig:cts-latent}
	\end{subfigure}
	\begin{subfigure}{.64\textwidth}
		 \centering
		 \includegraphics[width=.85\linewidth]{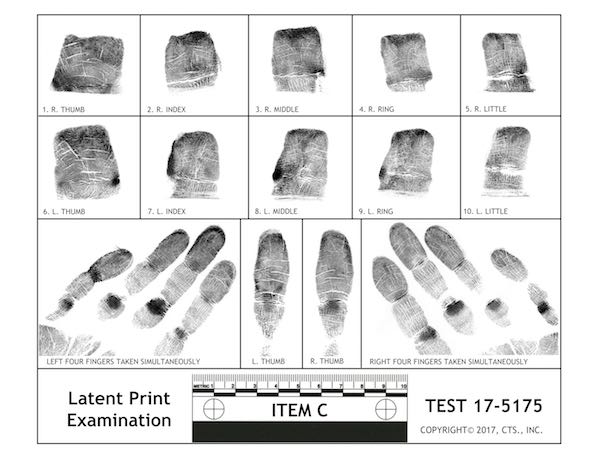}
		\caption{A ten-print card reference sample provided by CTS.}
		\label{fig:cts-reference}
	\end{subfigure}
	\label{fig:cts-example}
	\caption{Examples of latent and reference samples provided in CTS proficiency exams.}
\end{figure}

In a typical CTS exam, for example, 300--500 participants respond to eleven or twelve items.  In a typical item, a latent print is presented (e.g. Figure~\ref{fig:cts-latent}), and participants are asked to determine the source of the print from a pool of four known donors (e.g. Figure~\ref{fig:cts-reference}), if any.   

Proficiency tests may be used for training, known or blind proficiency testing, research and development of new techniques, etc.   Even non-forensic examiners can participate in CTS exams \citep{max2019assessing} and distinguishing between experts and non-experts from the response data alone is usually not feasible since most participants correctly answering every question  \citep{luby2018proficiency}. Moreover, since the test environment is not controlled, it is impossible to determine whether responses correspond to an individual examiner's decision, to the consensus answer of a group of examiners working together on the exam, or some other response process.  

\subsection{Error-rate Studies}\label{ss:error-rate-studies}

Error rate studies typically consist of a smaller
number of participants (fewer than \(200\)), but use a larger pool of items (often 100 or more). In general, the items are designed to be
difficult, and every participant does not respond to every item.

\citet{aaasReport} identified twelve existing error rate studies in the fingerprint domain, and a summary of those studies is provided here. The number of participants ($N$), number of items ($J$), false positive rate, false negative rate, and reporting strategy vary widely across the studies and are summarized in Table~\ref{studySummary} below.   For example, \citet{evett1996review} did report the number of inconclusive responses, making results difficult to evaluate relative to the other studies.  And  \citet{tangen2011identifying} and \citet{kellman2014} required examiners to make a determination about the source of a latent print in only three minutes, likely leading to larger error rates.  \citet{ulery2011accuracy} is generally regarded as the most well-designed error rate study for fingerprint examiners \citep{aaasReport,pcast}. \citet{ulery2012repeatability} tested the same examiners on 25 of the same items they were shown seven months earlier, and found that 90\% of decisions for same-source pairs were repeated, and 85.9\% of decisions for different-source pairs were repeated.  For additional information on all twelve studies, see \citet{luby2019thesis} or \citet{aaasReport}.  

\begin{table}[h]
	\centering
	\begin{tabular}{rrrrrr}
		\hline
		& $N$ & $J$ & False Pos & False Neg & Inconclusive \\ 
		\hline
		\citet{evett1996review} & 130 & 10  & 0 & 0.007\% & Not reported \\ 
		\citet{wertheim2006report} & 108 & 10 & 1.5\%  & & \\
		\citet{langenburgchampodwertheim2009testing}& 15 (43) & 6 & 2.3\%  & 7\% & \\ 
		\citet{langenberg2009performance}& 6 & 120 & 0 & 0.7\%/ 2.2\% & \\
		\citet{tangen2011identifying} & 37 (74) & 36 & 0.0037 &  & Not allowed \\
		\citet{ulery2011accuracy} & 169 & 744 (100)& 0.17\% & 7.5\% & \\
		\citet{ulery2012repeatability} & 72 & 744 (25) & 0 & 30\% of previous & \\
		\citet{langenburg2012informing} & 159 & 12 & 2.4\% &  & \\
		\citet{kellman2014} & 56 & 200 (40) & 3\% & 14\% & Not allowed \\
		\citet{pacheco2014miami} & 109 & 40 & 4.2\% & 8.7\%  & \\
		\citet{liu2015study} & 40 & 5 & 0.11\% & & \\
		\hline
	\end{tabular}
	\caption{Summary of existing studies that estimate error rates in fingerprint examination}
	\label{studySummary}
\end{table}

\subsection{FBI Black Box Study}\label{ss:black-box}

All analyses in this paper use results from the FBI Black Box Study and are based on practices and procedures of fingerprint examiners in the United States. The FBI Black Box study \citep[dataset available freely from the FBI\footnote{\url{https://www.fbi.gov/services/laboratory/scientific-analysis/coun\-ter\-ter\-ror\-ism-forensic-science-research/black-box-study-results}}]{ulery2011accuracy} was the first large-scale study performed to assess the accuracy and reliability of fingerprint examiners’ decisions. 169 fingerprint examiners were recruited for the study, and each participant was assigned roughly 100 items from a pool of 744. The items (fingerprint images) were designed to include ranges of features (e.g. minutiae, smudges, and patterns) and quality similar to those seen in casework, and to be representative of searches from an automated fingerprint identification system. The overall false positive rate in the study was 0.1\% and the overall false negative rate was 7.5\%. These computed quantities, however, excluded all ``inconclusive'' responses (i.e. neither individualizations nor exclusions).  

Each row in the data file corresponds to an examiner \(\times\) task response. In addition to the Examiner ID and item Pair ID (corresponding to the latent-reference pair), additional information is provided for each examinee $\times$ task interaction, as shown in Table~\ref{t:blackbox-fields}.

\begin{table}[ht]
\begin{center}
\begin{tabular}{|p{6in}|}
\hline
\begin{itemize}\tight
	\item
	\texttt{Mating}: whether the pair of prints were ``Mates'' (a match)
	or ``Non-mates'' (a non-match)
	\item
	\texttt{Latent\_Value}: the examiner's assessment of the value of the
	print (NV = No Value, VEO = Value for Exclusion Only, VID = Value for
	Individualization)
	\item
	\texttt{Compare\_Value}: the examiner's evaluation of whether the pair
	of prints is an ``Exclusion'', ``Inconclusive'' or
	``Individualization''
	\item
	\texttt{Inconclusive\_Reason}: If inconclusive, the reason for the
	inconclusive
	
	\begin{itemize}
		\item
		``Close'': \emph{The correspondence of features is supportive of the
			conclusion that the two impressions originated from the same source,
			but not to the extent sufficient for individualization.}
		\item
		``Insufficient'': \emph{Potentially corresponding areas are present,
			but there is insufficient information present.} Examiners were
		told to select this reason if the reference print was not of value.
		\item
		``No Overlap'': \emph{No overlapping area between the latent and
			reference prints}
	\end{itemize}
	\item
	\texttt{Exclusion\_Reason}: If exclusion, the reason for the exclusion
	
	\begin{itemize}
		\item
		``Minutiae'': \emph{The exclusion determination required the use of minutiae}
		\item
		``Pattern'':  \emph{The exclusion determination could be made on fingerprint pattern class and did not require the use of minutiae}
	\end{itemize}
	\item
	\texttt{Difficulty}: Reported difficulty on a five point scale: `A-Obvious', `B-Easy', `C-Medium', `D-Difficult', `E-Very Difficult'.
\end{itemize} \\
\hline
\end{tabular}
\end{center}
\caption{Additional information provided for each examiner $\times$ task interaction in the FBI Black Box data \citep{ulery2011accuracy}.}\label{t:blackbox-fields}
\end{table}

Examiners thus made three distinct decisions when they were evaluating the latent and reference prints in each item: (1) whether or not the latent print has value for a further decision, (2) whether the latent print was determined to come from the same source as the reference print, different sources, or inconclusive, and (3) their reasoning for making an inconclusive or exclusion decision. While the main purpose of the study was to calculate casework error rates (and thus focused on the \texttt{Compare\_Value} decision), important trends in examiner behavior are also present in the other decisions, to which we return in Section \ref{ss:structured-responses}. 

\section{Proficiency and Process Modeling for Fingerprint Examiners}\label{s:modeling}

\subsection{Applying the Rasch model}
\label{ss:rasch}

The Rasch Model \citep{rasch1960studies, raschbook} is a relatively
simple, yet powerful, item response model, that allows us to separate examiner proficiency from task difficulty. The probability of a correct response is
modeled as a logistic function of the difference between the participant
proficiency, \(\theta_i\) (\(i=1, \dots, N\)), and the item difficulty,
\(b_j\) (\(j=1, \dots, J\)),
\begin{equation}
P(Y_{ij} = 1) = \frac{1}{1-\exp(-(\theta_i - b_j))}. 
\label{rasch}
\end{equation}
%

In order to fit an IRT model to the Black Box Study, we will score responses as correct if they are true identifications
or exclusions
and as incorrect if they are false identifications
or exclusions.
%
For the purpose of illustration will consider ``inconclusive'' responses as missing completely at random (MCAR), following \citet{ulery2011accuracy}.  However, there are a large number of inconclusive answers (4907 of 17121 responses), which can be scored in a variety of ways \citep[see][for examples]{luby2019openforsci}, and we will return to the inconclusives in Section~\ref{ss:no-answer-key}. 

\begin{figure}
	\centering
	\includegraphics[width=0.7\linewidth]{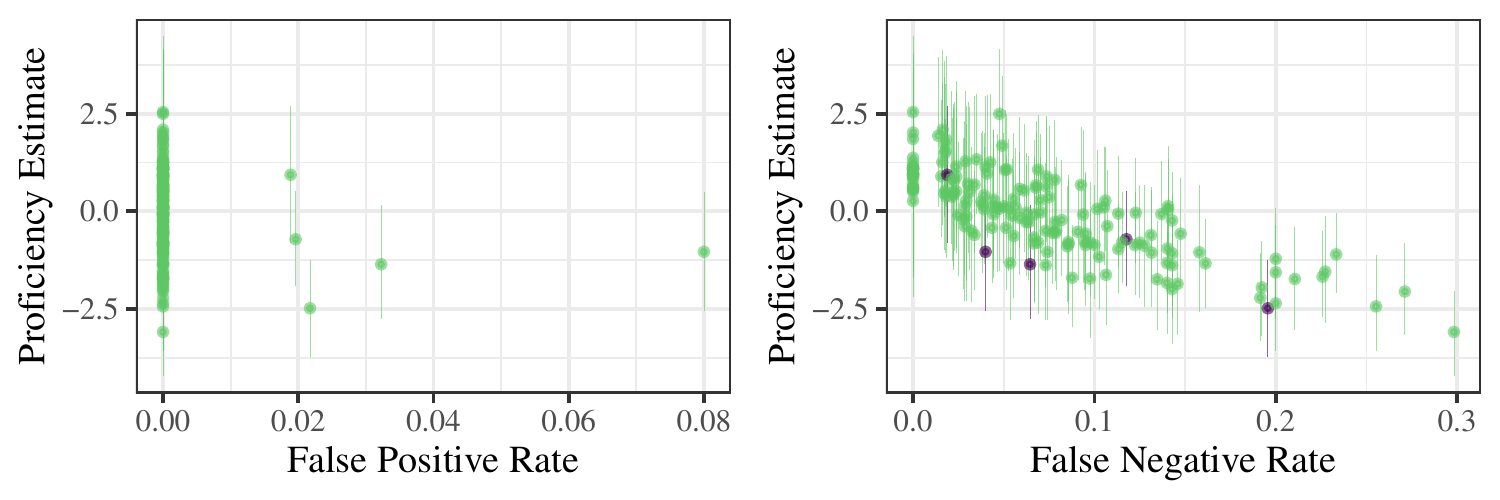}
	\caption{Estimated IRT proficiency by observed false positive rate (left panel) and false negative rate (right panel). Examiners who made at least one false positive error, i.e. the nonzero cases in the left-hand plot, are colored in purple on the right-hand plot.}
	\label{fig:prof-error-rate}
\end{figure}

The Rasch model was fitted in a Bayesian framework, with $\theta_i \sim N(0, \sigma_\theta^2)$, $b_j \sim N(\mu_b, \sigma_b^2)$, $\mu_b \sim N(0,10)$, $\sigma_\theta \sim \text{Half-Cauchy}(0, 2.5)$ and $\sigma_b \sim \text{Half-Cauchy}(0, 2.5)$, using Stan \citep{rstan, stan}. Figure \ref{fig:prof-error-rate} shows estimated proficiencies of examiners when responses are scored as described above, with 95\% posterior intervals, plotted against the raw false positive rate (left panel) and against the raw false negative rate (right panel). Those examiners who made at least one false positive error are colored in purple in the right panel of Figure~\ref{fig:prof-error-rate}. One of the examiners who made a false positive error still received a relatively high proficiency estimate due to having a small false negative rate. 

\begin{figure}
	\centering
	\includegraphics[width=0.7\linewidth]{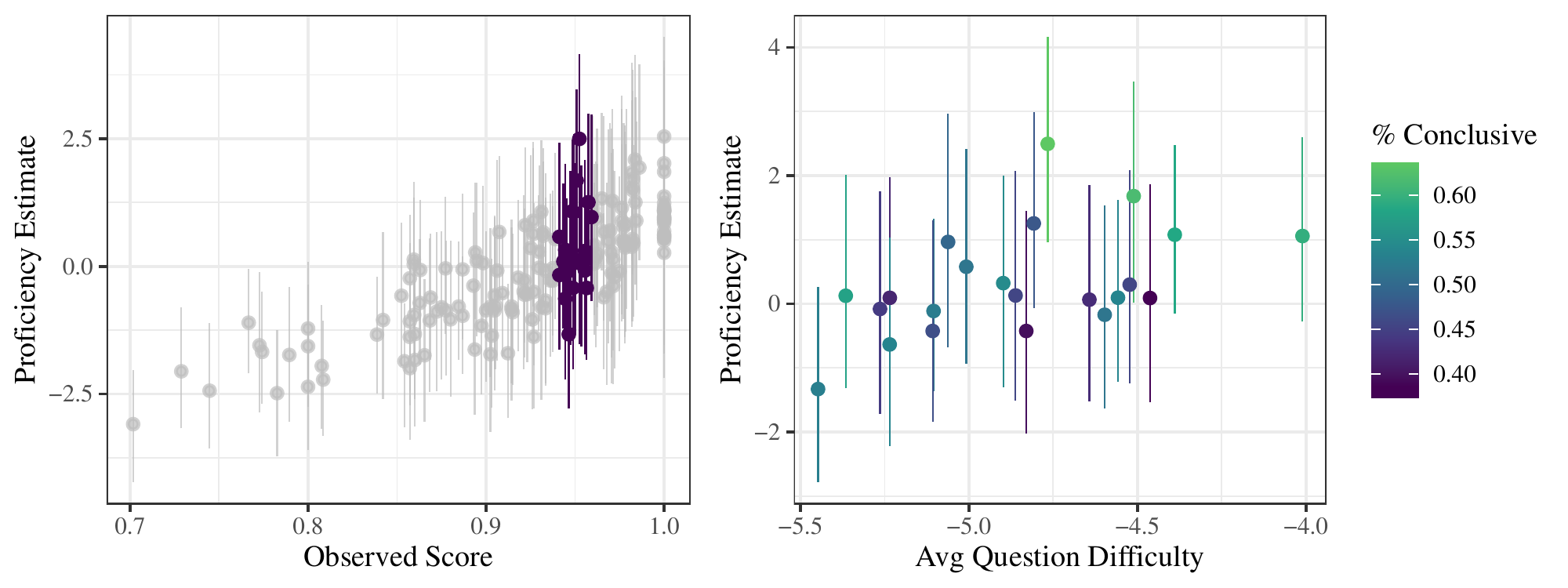}
	\caption{The left panel shows proficiency by observed score under the ``inconclusive MCAR'' scoring scheme, with those examiners with scores between 94\% and 96\% highlighted. The right panel shows proficiency by average item difficulty, colored by percent conclusive, for the highlighted subset from the left panel. Estimated proficiency is related to observed score, item difficulty, and conclusive decision rates.}
	\label{fig:prof-observed}
\end{figure}

In the left panel of Figure~\ref{fig:prof-observed}, we see as expected a positive correlation between proficiency estimates and observed score (\% correct); variation in proficiency at each observed score is due to the fact that different examiners saw subsets of items of differing difficulty.   The highlighted examiners in the left panel in Figure~\ref{fig:prof-observed} all had raw percent-correct (observed scores) between 94\% and 96\%, and are re-plotted in the right panel showing average question difficulty, and percent of items with conclusive responses, illustrating substantial variation in both Rasch proficiency and relative frequency of conclusive responses, for these examiners with similar, high observed scores.

\cite{luby2019openforsci} explores other scoring schemes as well as partial credit models for this data.  Treating the inconclusives as MCAR leads to both the smallest range of observed scores and largest range of estimated proficiencies; harsher scoring methods (e.g. treating inconclusives as incorrect) generally also lead to higher estimated proficiencies, since more items are estimated to be difficult.  

Results from an IRT analysis are largely consistent with conclusions from an error rate analysis \citep{luby2019openforsci}. However, IRT provides substantially more information than a more traditional analysis, specifically through accounting for the difficulty of items seen. Additionally, IRT implicitly accounts for the inconclusive rates of different examiners in its estimates of uncertainty for both examiner proficiency and item difficulty.

\subsection{Covarying Responses: Self-reported Difficulty}
\label{ss:covarying-responses}

As shown in Table~\ref{t:blackbox-fields}, the FBI Black Box study also asked examiners to report the difficulty of each item they evaluated on a five-point scale. These reported difficulties are not the purpose of the test, but are secondary responses for each item collected at the same time as the responses and can therefore be thought of as `collateral information'. 
When the additional variables are covariates describing either the items or the examiners---for instance, image quality, number of minutiae, examiner's experience, type of training---it would be natural to incorporate them as predictors for proficiency or difficulty in the IRT model \citep{eirtbook}.  However, since reported difficulty is, in effect, a secondary response in the Black Box study, we take an approach analogous to response time modeling in IRT: in our case we have a scored task response, and a difficulty rating rather than a response time, for each person $\times$ item pair.  \citet{thissen_9_1983} provides an early example of this type of modeling, where the logarithm of response time is modeled as a linear function of the log-odds $\theta_i - b_j$ of a correct response, and additional latent variables for both items and participants. \citet{ferrando_item_2007} and  \citet{van_der_linden_lognormal_2006}  each propose various other models for modeling response time jointly with the traditional correct/incorrect IRT response. Modeling collateral information alongside responses in this way has been shown generally to improve estimates of IRT parameters through the sharing of information \citep{van2010irt}. 

\subsubsection{Model}\label{ss:reported-difficulty-model}

Recall from Section~\ref{ss:black-box} (Table~\ref{t:blackbox-fields}) that examiners rate the difficulty of each item on a five-point scale: `A-Obvious', `B-Easy', `C-Medium', `D-Difficult', `E-Very Difficult'.  
Let $Y_{ij}$ be the scored response of participant $i$ to item $j$, and let $X_{ij}$ be the difficulty reported by participant $i$ to item $j$. $Y_{ij}$ thus takes the values 0 (incorrect) or 1 (correct), and $X_{ij}$ is an ordered categorical variable with five levels (A-Obvious to E-Very Difficult).
Following \citet{thissen_9_1983},
we combine a Rasch model,
\begin{equation}
\logit(P(Y_{ij} = 1)) = \theta_i  - b_j ,
\end{equation}
with a cumulative-logits ordered logistic model for the reported difficulties,
\begin{equation}
X^*_{ij} = \logit^{-1} (g \cdot (\theta_i - b_j) + h_i + f_j) ,
\label{eq:thissen-part}
\end{equation}
where 
\begin{equation}
X_{ij} = \begin{cases}
\text{A-Obvious} & X^*_{ij} \le \gamma_1 \\
\text{B-Easy} & \gamma_1 < X^*_{ij} \le \gamma_2 \\
\text{C-Medium} & \gamma_2 < X^*_{ij} \le \gamma_3 \\
\text{D-Difficult} & \gamma_3 < X^*_{ij} \le \gamma_4 \\
\text{E-Very Difficult} & X^*_{ij} > \gamma_4.
\end{cases}
\label{eq:rep-diff-mod}
\end{equation}

\noindent
The additional variables $h_i$ and $f_j$ in equation (\ref{eq:thissen-part}) allow for the possibilities that examiners over-report ($h_i>0$) or under-report ($h_i<0$) item difficulty, and that item difficulty tends to be over-reported ($f_j>0$) or under-reported ($f_j<0$), relative to the Rasch logit $(\theta_i - \beta_j)$ and the reporting tendencies of other examiners.  These parameters will be discussed further in Section~\ref{ss:reported-difficulty-results} below.

We assume that each participant's responses are independent of other participants' responses, $Y_{i\cdot} \perp Y_{i'\cdot}$; that within-participant responses and reports are conditionally independent of one another given the latent trait(s), $Y_{ij} \perp Y_{ij'} | \theta_i$ and $X_{ij} \perp X_{ij'} | \theta_i, h_i$; and that responses are conditionally independent of reported difficulty given all latent variables, $X_{ij} \perp Y_{ij} | \theta_i, b_j, g, h_i, f_j$. Then the likelihood is
\begin{equation}
L(Y,X | \theta, b, g, h_i, f_j) = \prod_{i} \prod_{j} P(Y_{ij} = 1)^{Y_{ij}}  (1-P(Y_{ij} = 1))^{1-Y_{ij}} P(X_{ij} = x_{ij})
\end{equation}
and 
\begin{equation}
P(X_{ij} = c) = P(\logit^{-1} (g \cdot (\theta_i - b_j) + h_i + f_j) \le \gamma_c) - P(\logit^{-1} (g \cdot (\theta_i - b_j) + h_i + f_j) \le \gamma_{c-1}),
\end{equation}
where $\gamma_0 = -\infty$ and $\gamma_5 = \infty$. 

We chose a cumulative-logits approach because it is directly implemented in Stan and therefore runs slightly faster than adjacent-category-logits and other approaches. We have no reason to believe this choice has a practical effect on modeling outcomes, but if desired other formulations could certainly be used. \citet{luby2019thesis} compares the predictive performance and prediction error of the above model with fits of other models for $X_{ij}$ and finds the above model to best fit the Black Box data.

\subsubsection{Results}\label{ss:reported-difficulty-results}

For each examiner in the dataset, their observed score
$\frac{1}{n_i} \sum_{j \in J_i} y_{ij}$, and their predicted score under the model, $\frac{1}{n_i} \sum_{j \in J_i} \hat{y}_{ij}$, were calculated. Similarly, predicted and observed average reported difficulty were calculated, where the observed average reported difficulty is $\frac{1}{n_i} \sum_{j \in J_i} x_{ij}$ and the predicted average reported difficulty is $\frac{1}{n_i} \sum_{j \in J_i} \hat{x}_{ij}$. If the model is performing well, the predicted scores should be very similar to the observed scores. 

\begin{figure}
	\begin{minipage}{.47\linewidth}
		\includegraphics[width=\linewidth]{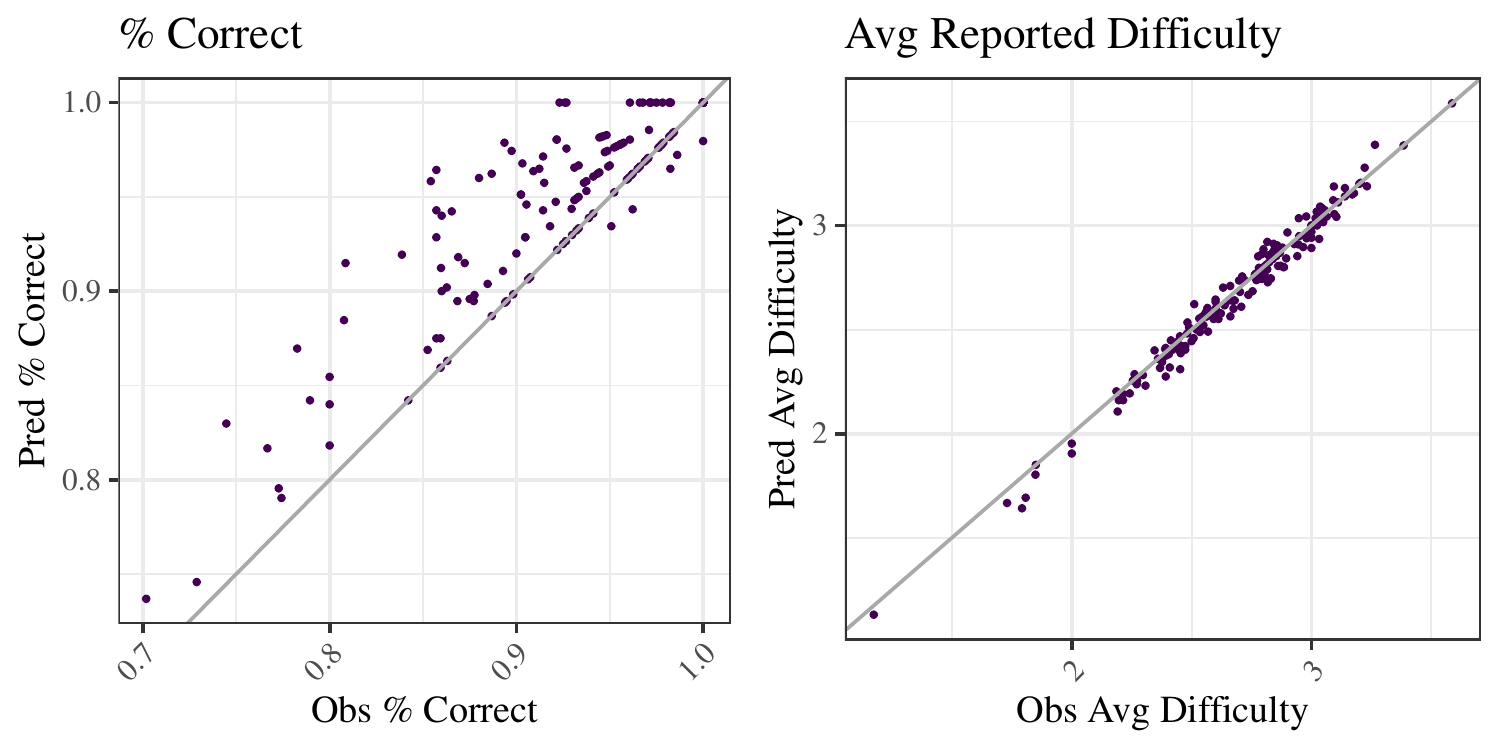}
		\caption{Posterior predictive performance of \% correct (left) and average predicted difficulty (right) for the joint model. The model slightly over-predicts \% correct, but performs quite well for average reported difficulty.}
		\label{fig:pp-thissen}
	\end{minipage}
	\hfill
	\begin{minipage}{.47\linewidth}
		\includegraphics[width=\linewidth]{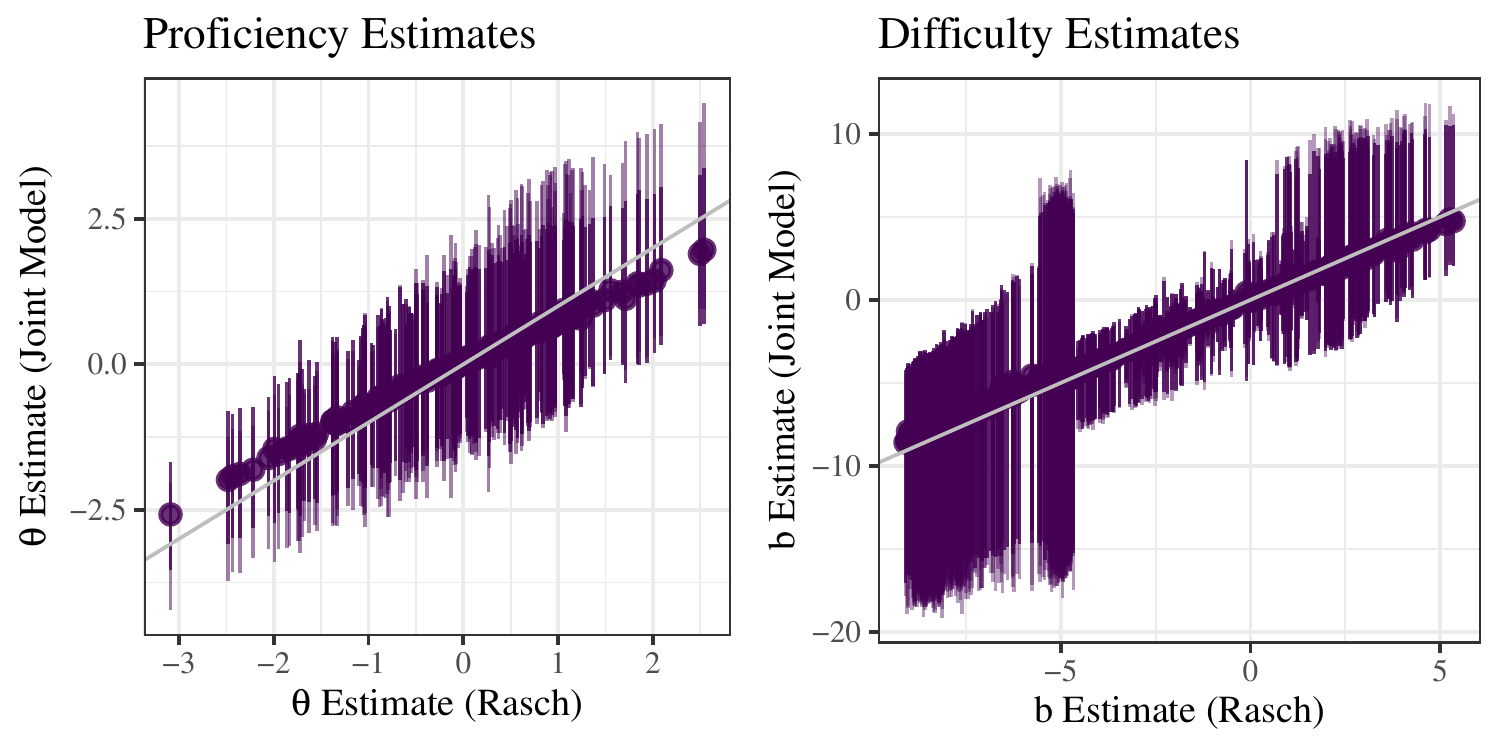}
		\caption{Proficiency (left) and difficulty (right) estimates under the joint model (with 95\% posterior intervals) are very similar to Rasch proficiency point estimates from previous section.}
		\label{fig:prof-comparison-rasch-thissen}
	\end{minipage}
\end{figure}

Figure~\ref{fig:pp-thissen} shows the predicted scores compared to the observed scores (left panel), and the predicted average difficulty compared to the observed average reported difficulty (right panel).  Reported difficulties for inconclusive responses were also treated as MCAR under this scoring scheme. While the joint model tends to over-predict percent correct, it predicts average reported difficulty quite well.

Figure~\ref{fig:prof-comparison-rasch-thissen} (left panel) plots the proficiency estimates from the joint model against the Rasch proficiency estimates (i.e.\ the model for correctness from Section~\ref{ss:rasch} \textit{without} modeling reported difficulty). The proficiency estimates from the joint model do not differ substantially from the Rasch proficiency estimates, although there is a slight shrinkage towards zero of the joint model proficiency estimates. Figure~\ref{fig:prof-comparison-rasch-thissen} (right panel) plots the item difficulty estimates from the joint model against the item difficulty estimates from the Rasch model. Like proficiency estimates, the difficulties under the joint model do not differ substantially from the Rasch difficulties. This is due to the inclusion of the $h_i$ and $f_j$ parameters for the reported difficulty part of the model, which sufficiently explains the variation in reported difficulty without impacting the IRT parameters. 

Recall that the joint model predicts reported difficulty as $g \cdot (\theta_i - b_j) + h_i + f_j$. In addition to proficiency and difficulty, ``reporting bias" parameters for examiners ($h_i$) and items ($f_j$) are also included. Positive $h_i$ and $f_j$ thus \textit{increase} the expected reported difficulty while negative $h_i$ and $f_j$ \textit{decrease} the expected reported difficulty. 

\begin{figure}[h]
	\centering
	\includegraphics[width=0.7\linewidth]{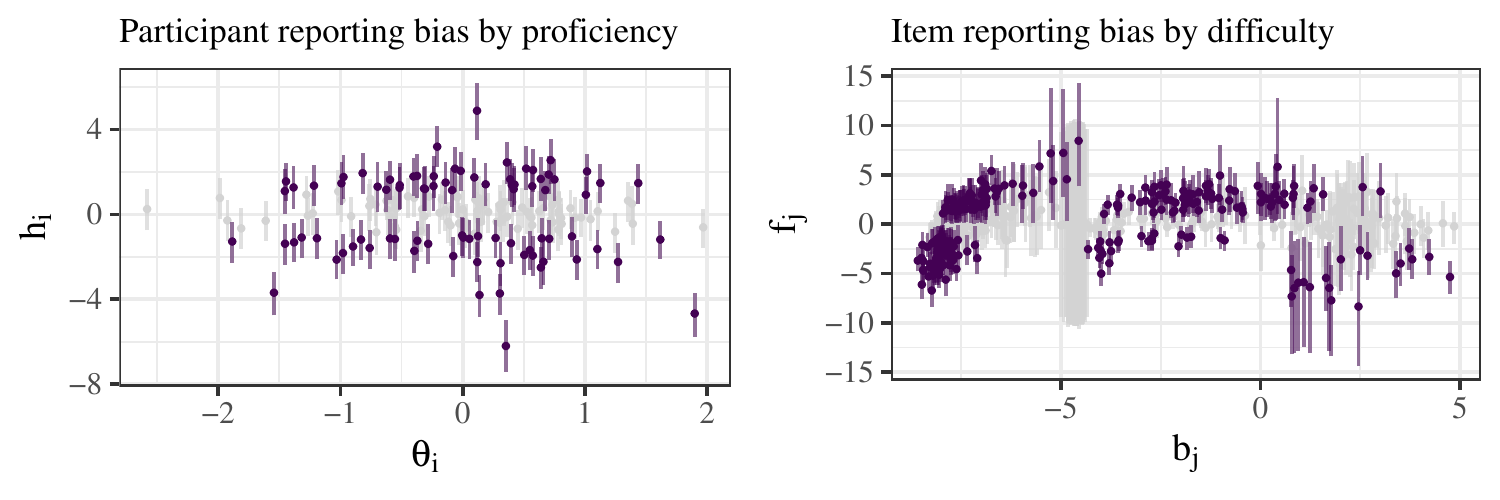}
	\caption{Person reporting bias ($h_i$, left) and item reporting bias ($f_j$, right) with 95\% posterior intervals from the Thissen model compared to proficiency estimate ($\theta_i$) and difficulty estimate ($b_j$), respectively. Points with intervals that overlap with zero are colored in gray. There is substantial variation in $h_i$ not explained by $\theta_i$. Items with estimated difficulties near zero are most likely to have over-reported difficulty.}
	\label{fig:reporting-bias}
\end{figure}

Thus, $h_i$ can be interpreted as examiner $i$'s tendency to over or under-report difficulty, after accounting for the other parameters. The left panel of Figure~\ref{fig:reporting-bias} shows the $h_i$ estimates and 95\% posterior intervals compared to the proficiency (point) estimates. Since there are many examiners whose 95\% posterior intervals do not overlap with zero, Figure~\ref{fig:reporting-bias} provides evidence that there exist differences among examiners in the way they report difficulty. This reporting bias does not appear to have any relationship with the model-based proficiency estimates. That is, examiners who report items to be more difficult (positive $h_i$) do not perform worse than examiners who report items to be easier (negative $h_i$). 

Similarly, $f_j$ can be interpreted as item $j$'s tendency to be over or under-reported, after accounting for other parameters. The right panel of Figure~\ref{fig:reporting-bias} shows the $f_j$ estimates and 95\% posterior intervals compared to the point estimates for difficulty ($b_j$).  There are a substantial number of items whose posterior intervals do not overlap with zero; these are items that are consistently reported as more or less difficult than the number of incorrect responses for that item suggests. Additionally, there is a mild arc-shaped relationship between $f_j$ and $b_j$: items with estimated difficulties near zero are most likely to have over-reported difficulty, and items with very negative or very positive estimated difficulties (corresponding to items that examiners did very poorly or very well on, respectively) tend to have under-reported difficulty. 

Reported difficulty may provide additional information about the items beyond standard IRT estimates. For example, consider two items with identical response patterns (i.e. the same examiners answered each question correctly and incorrectly) but one item was reported to be more difficult than the other by all examiners. It is plausible that at least some examiners struggled with that item, but eventually came to the correct conclusion.  Standard IRT will not detect the additional effort required for that item, compared to the less effortful item with the same response pattern.  

\subsection{Sequential Responses}
\label{ss:structured-responses}

Although the purpose of the Black Box study was to estimate false positive and false negative error rates, the recorded data also contains additional information about examiners' decision-making process. Recall from Section~\ref{ss:black-box} that each recorded response to an item consists of three decisions: 

\begin{enumerate}
	\item Value assessment for the latent print only  (No Value, Value for Exclusion Only, or Value for Individualization)
	\item Source evaluation of the latent/reference print pair (i.e. Individualization [match], Exclusion [non-match], or Inconclusive)
	\item (If inconclusive) Reason for inconclusive
\end{enumerate}

For our analysis, we do not distinguish between `value for individualization' and `value for exclusion only', and instead treat the value assessment as a binary response (`Has value' vs `No value').  As \citet{haber2014experimental} note, only 17\% of examiners reported that they used `value for exclusion only' in their normal casework on a post-experiment questionnaire, and examiners in the Black Box study may have interpreted this decision in different ways. For example, there were 32 examiners (of 169) who reported that a latent print had `value for exclusion only' and then proceeded to make an individualization for the second decision.  These discrepancies led us to treat the value evaluation as a binary response -- either `has value' or `no value'. 

The Item Response Trees (IRTrees, \citealp{deboeck2012statsoft}) framework provides a solution for modeling the sequential decisions above explicitly. IRTrees represent responses with decision trees where branch splits represent hypothesized internal decisions, conditional on the previous decisions in the tree structure, and leaves are observed outcomes. Sequential decisions can be represented explicitly in the IRTree framework, and node splits need not represent scored decisions.

\begin{figure}
	\centering
	\includegraphics[width=0.7\linewidth]{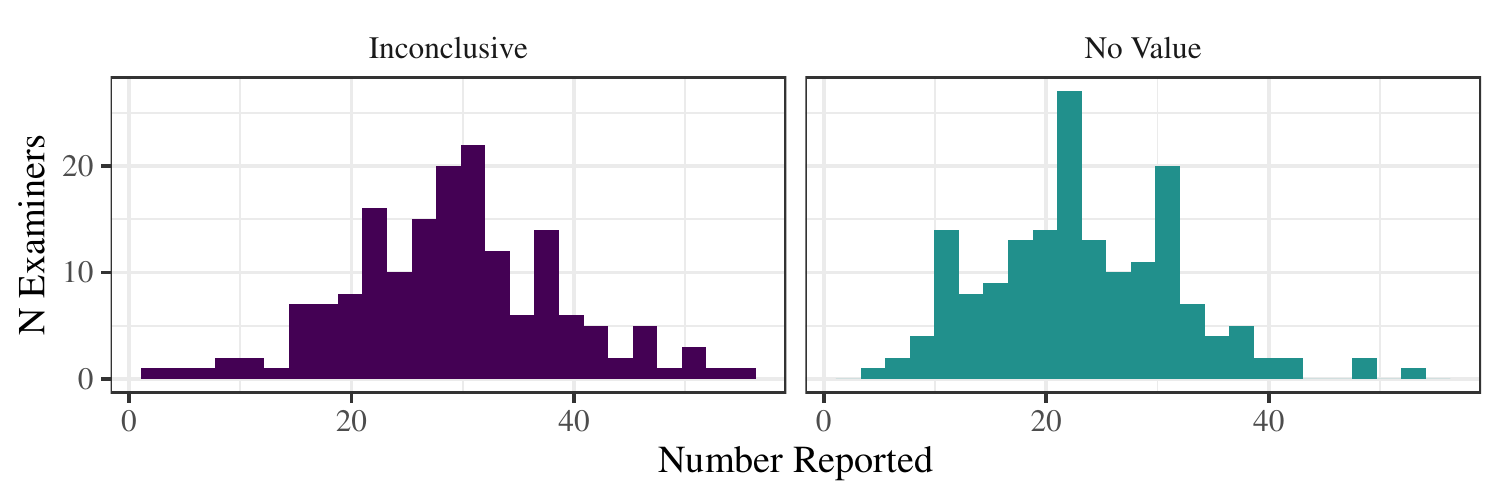}
	\caption{Number of inconclusive (left) and no value (right) responses reported by each examiner.}
	\label{fig:inc-nv-hists}
\end{figure}

Fingerprint examiners have been found to vary in their tendencies to make `no-value' and  `inconclusive' decisions \citep{ulery2011accuracy}. Figure~\ref{fig:inc-nv-hists} shows the distribution of the number of inconclusive and no value decisions reported by each examiner. Although most examiners report 20--40 inconclusives and 15--35 `no value' responses, some examiners report as much as 60 or as few as 5. By modeling these responses explicitly within the IRTree framework, individual differences in proficiency among examiners be assessed alongside differences in tendency towards value assessments (vs no-value assessments) and inconclusive responses (vs conclusive responses). 

\subsubsection{Model}

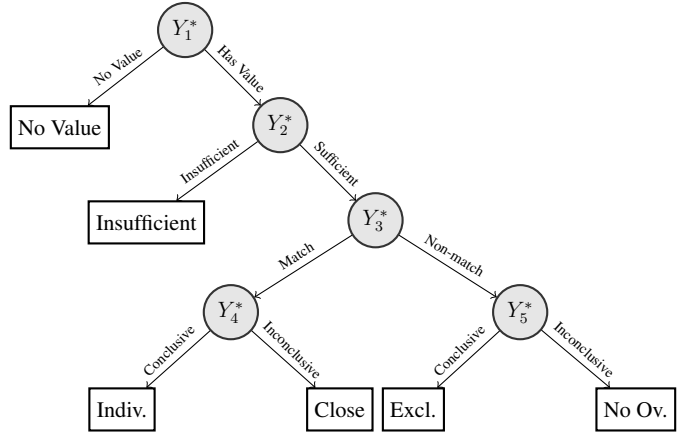
\begin{figure}
	\centering
	\begin{tikzpicture}[scale = .7, transform shape]
	\tikzstyle{latent}=[circle, minimum size = 9mm, thick, draw =black!80, node distance = 15mm, fill = black!10]
	\tikzstyle{observed}=[rectangle, minimum size = 8mm, thick, draw = black!100, node distance = 15mm]
	\node[latent] (y0) {$Y^*_{1}$};
	\node[observed] (nv) [below left = of y0] {No Value};
	\node[latent] (y1) [below right = of y0]{$Y^*_{2}$};
	\node[latent] (y2) [below right = of y1]{ $Y^*_{3}$ };
	\node[observed] (in) [ below left = of y1] {Insufficient};
	\node[latent] (y3) [below left = 10mm and 20mm of y2] { $Y^*_{4}$ };
	\node[latent] (y3b) [below right = 10mm and 20mm of y2] { $Y^*_{5}$ };
	\node[observed] (id) [below left=of y3]{Indiv.};
	\node[observed] (in1) [below right =of y3] {Close};
	\node[observed] (ex) [below left =of y3b] {Excl.};
	\node[observed] (in2) [below right = of y3b] {No Ov.};
	\draw[->] (y0) -> (y1) node [midway, rotate=-45, anchor = south]  {\scriptsize Has Value};
	\draw[->] (y0) -> (nv) node [midway, rotate=35, anchor = south]  {\scriptsize No Value};
	\draw[->] (y1) -> (y2) node [midway, rotate=-45, anchor = south]  {\scriptsize Sufficient};
	\draw[->] (y1) -> (in) node [midway, rotate=35, anchor = south]  {\scriptsize Insufficient};
	\draw[->] (y2) -> (y3) node [midway, rotate=32, anchor = south]  {\scriptsize Match};
	\draw[->] (y2) -> (y3b) node [midway, rotate=-32, anchor = south]  {\scriptsize Non-match};
	\draw[->] (y3) -> (id) node [midway, rotate=45, anchor = south]  {\scriptsize Conclusive};
	\draw[->] (y3) -> (in1) node [midway, rotate=-45, anchor = south]  {\scriptsize Inconclusive};
	\draw[->] (y3b) -> (ex) node [midway, rotate=45, anchor = south]  {\scriptsize Conclusive};
	\draw[->] (y3b) -> (in2) node [midway, rotate=-42, anchor = south]  {\scriptsize Inconclusive};
	\end{tikzpicture}
	\caption{The \textit{binary decision process} tree}
	\label{hyp-process-tree}
\end{figure}

Figure~\ref{hyp-process-tree} depicts an IRTree based on one possible internal decision process, motivated by the ACE-V decision process \citepalias{nist2012latent}.  Each internal node $Y^*_1, \ldots, Y^*_5$ represents a possible binary (0/1) decision that each examiner could makes on each item; these decisions will be modeled with IRT models.  The first node, $Y_{1}^*$, represents the examiner's assessment of whether the latent print is ``of value'' or ``no value''. The second node, $Y_{2}^*$ represents whether the examiner found sufficient information in the (reference, latent) print pair to make a further decision. $Y_{3}^*$ represents whether the pair of prints is more likely to be a match or a non-match, and $Y_{4}^*$ and $Y_{5}^*$ represent whether this determination is conclusive (individualization and exclusion, respectively) or inconclusive (close and no overlap, respectively). This binary decision process tree thus separates examiners' decisions into both (a) distinguishing between matches and non-matches ($Y_{3}^*$) and (b) examiner ``willingness to respond with certainty'' ($Y_{1}^*, Y_2^*, Y_4^*, Y_5^*$). 

Since each internal node in the IRTree in Figure~\ref{hyp-process-tree} is a binary split, we use a Rasch model to parameterize each branch in the tree. That is, 
\begin{equation}
P(Y_{kij}^* = 1) = \text{logit}^{-1}(\theta_{ki} - b_{kj}),
\label{eq:IRTrees-rasch}
\end{equation}
where $i$ indexes examiners, $j$ indexes items, and $k$ indexes internal nodes (sequential binary decisions). Thus, we account for examiner tendencies to choose one branch vs. the other at decision $k$ with $\theta_{ki}$, and features of the task that encourage choice of one branch vs. the other at decision $k$ with $b_{kj}$. Clearly other IRT models could be chosen as well; see \citet{luby2019thesis} for further discussion.  The full IRTree model is 
\begin{align}
P(Y_{ij}= \text{No Value})  &= P(Y_{1ij}^* = 1) \\
P(Y_{ij}= \text{Individ.})  &= P(Y_{1ij}^* = 0) \times P(Y_{2ij}^* = 0) \times P(Y_{3ij}^* = 1) \times P(Y_{4ij}^* = 1)\\
P(Y_{ij}= \text{Close})  &= P(Y_{1ij}^* = 0) \times P(Y_{2ij}^* = 0) \times P(Y_{3ij}^* = 1) \times P(Y_{4ij}^* = 0) \\
P(Y_{ij}= \text{Insufficient})  &= P(Y_{1ij}^* = 0) \times P(Y_{2ij}^* = 1) \\
P(Y_{ij}= \text{No Ov.})  &= P(Y_{1ij}^* = 0) \times P(Y_{2ij}^* = 0) \times P(Y_{3ij}^* = 0) \times P(Y_{5ij}^* = 0) \\
P(Y_{ij}= \text{Excl.})  &= P(Y_{1ij}^* = 0) \times P(Y_{2ij}^* = 0) \times P(Y_{3ij}^* = 0) \times P(Y_{5ij}^* = 1)  .
\end{align}
Furthermore, an item-explanatory variable ($X_j$) for each item was included at all $k$ nodes, where $X_j = 1$ if the latent and reference print came from the same source (i.e. a true match) and $X_j = 0$ if the latent and reference print came from different sources (i.e. a true non-match). Then, 
\begin{equation}
b_{kj} = \beta_{0k} + \beta_{1k} X_{j} + \epsilon_{jk} \hspace{1cm} k=1,...,5, \label{lltmequation}
\end{equation}
where $b_{kj}$ are the item parameters and $\beta_{0k}, \beta_{1k}$ are linear regression coefficients at node $k$. This is an instance of the Linear Logistic Test Model \citep{fischer1973linear} with random item effects \citep{janssen2004models}; see also \citet{eirtbook} for more elaborate models along these lines. This allows for the means of item parameters to differ depending on whether the pair of prints is a true match or not.  The random effects $\epsilon_{kj} \sim N(0,\sigma^2_{kb})$, as specified in the second line of (\ref{eq:irtree-priors}) below, allow for the possibility that print pairs in an identification task may have other characteristics that impact task difficulty (e.g. image quality, number of features present), beyond whether the pair of prints is a same-source or different-source pair.

We fit this model under the Bayesian framework with Stan in R \citep{rstan, Rman}, using the following prior distributions,
\begin{equation}
\left.
	\begin{array}{r@{\,\,}l}
	\boldsymbol{\theta}_i &\stackrel{iid}{\sim} MVN_5(\boldsymbol{0}, \boldsymbol{\sigma_\theta }L_\theta L_\theta'\boldsymbol{\sigma_\theta }) \\
	\boldsymbol{b}_j &\stackrel{iid}{\sim} MVN_5(\boldsymbol{\beta}{\cal X}_j, \boldsymbol{\sigma_b}L_b L_b'\boldsymbol{\sigma_b }) \\
	L_\theta &\sim LKJ(4) \\
	L_{b} &\sim LKJ(4)\\
	\sigma_{k\theta} &\stackrel{iid}{\sim} \text{Half-Cauchy}(0, 2.5) \hspace{1cm} k=1,...,5\\
	\sigma_{kb} &\stackrel{iid}{\sim} \text{Half-Cauchy}(0, 2.5) \hspace{1cm} k=1,...,5\\
	\beta_{0k} &\stackrel{iid}{\sim} N(0,1) \hspace{1cm} k=1,...,5\\
	\beta_{1k} &\stackrel{iid}{\sim} N(0,1) \hspace{1cm} k=1,...,5.\\
	\end{array}
\hspace{1cm}\right\}\label{eq:irtree-priors}
\end{equation}
Here ${\cal X}_j$ is the column vector $(1, X_j)'$, 
$\boldsymbol{\beta} = (\boldsymbol{\beta_1}, ..., \boldsymbol{\beta_5})$ is the $5\times2$ matrix whose $k^{th}$ row is $(\beta_{0k}, \beta_{1k})$,  and $\boldsymbol{\sigma_b}$ is a $5\times5$ diagonal matrix with $\sigma_{1b}, ..., \sigma_{5b}$ as the diagonal entries; $\boldsymbol{\sigma_\theta}$ in the previous line is defined similarly. Multivariate normal distributions for $\boldsymbol{\theta}_i$ and $\boldsymbol{b}_j$ were chosen to estimate covariance between sequential decisions explicitly. The Stan modeling language does not rely on conjugacy, so the Cholesky factorizations ($L_\theta$ and $L_b$) are modeled instead of the covariance matrices for computational efficiency. The recommended priors \citep{stan} for $L$ and $\sigma$ were used: an LKJ prior \citep[LKJ = last initials of authors]{lkjprior} with shape parameter 4, which results in correlation matrices that mildly concentrate around the identity matrix ($LKJ(1)$ results in uniformly sampled correlation matrices), and half-Cauchy priors on $\sigma_{kb}$ and $\sigma_{k\theta}$ to weakly inform the correlations. $N(0,5)$ priors were assigned to the linear regression coefficients ($\beta_k$). 

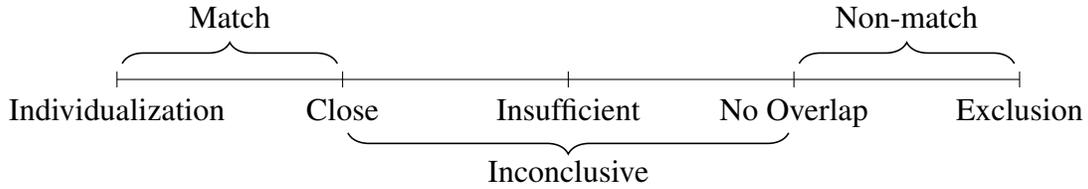
\begin{figure}[h]
	\centering
	\begin{tikzpicture}[Brace/.style args = {#1}{semithick, decorate, decoration={brace,#1,raise=6pt, amplitude = 8pt,
			pre=moveto,pre length=2pt,post=moveto,post length=2pt,}}]
	\draw (0,0) -- (12,0);
	\foreach \x in {0,3,6,9,12}
	\draw (\x cm,3pt) -- (\x cm,-3pt);
	
	\draw (0,0) node[below=3pt] {Individualization} node[above=3pt] {};
	\draw[Brace] (0,0) -- node[above=15pt] {Match} (3,0);
	\draw (3,0) node[below=3pt] {Close} node[above=3pt] {};
	\draw (6,0) node[below=3pt] {Insufficient} node[above=3pt] {};
	\draw[Brace=mirror] (3,-.5) -- node[below=12pt] {Inconclusive} (9,-.5);
	\draw (9,0) node[below=3pt] {No Overlap} node[above=3pt] {};
	\draw[Brace] (9,0) -- node[above=15pt] {Non-match} (12,0);
	\draw (12,0) node[below=3pt] {Exclusion} node[above=3pt] {};
	\end{tikzpicture}
	\caption{FBI black box responses as a bipolar scale. } 
	\label{bipolar-scale}
\end{figure}

There are, of course, alternative prior structures, and indeed alternate tree formulations, that could reasonably model this data.
For example \citet{luby2019thesis} constructs a novel bipolar scale, shown in Figure~\ref{bipolar-scale}, from the possible responses, and a corresponding IRTree model. This not only provides an ordering for the responses within each sub-decision (i.e. source decision and reason for inconclusive), but allows the sub-decisions to be combined in a logical way. This scale is also consistent with other hypothetical models for forensic decision-making \citep{dror2019cannot}. Based on the description of each option for an inconclusive response, the `Close' inconclusives are more similar to an individualization than the other inconclusive reasons. The `No overlap' inconclusives are more similar to exclusions than the other inconclusive reasons, under the assumption that the reference prints are relatively complete. That is, if there are no overlapping areas between a latent print and a complete reference print, the two prints likely came from different sources. The `insufficient' inconclusives are treated as the center of the constructed match/no-match scale. For more details, and comparsions among multiple tree structures, see \citet{luby2019thesis}. 

\subsubsection{Results}\label{ss:sequential-responses-results}

Our discussion of results will focus on estimated parameters from the fitted IRTree model.  For brevity, we will write $\theta_k = (\theta_{k1}, \ldots, \theta_{kN})$ and $b_k = (b_{k1}, \ldots, b_{kJ})$, $k=1, \ldots, 5$, in equation (\ref{eq:IRTrees-rasch}) and Figure~\ref{hyp-process-tree}.  

\begin{figure}[ht]
		\centering
		\includegraphics[width=\linewidth]{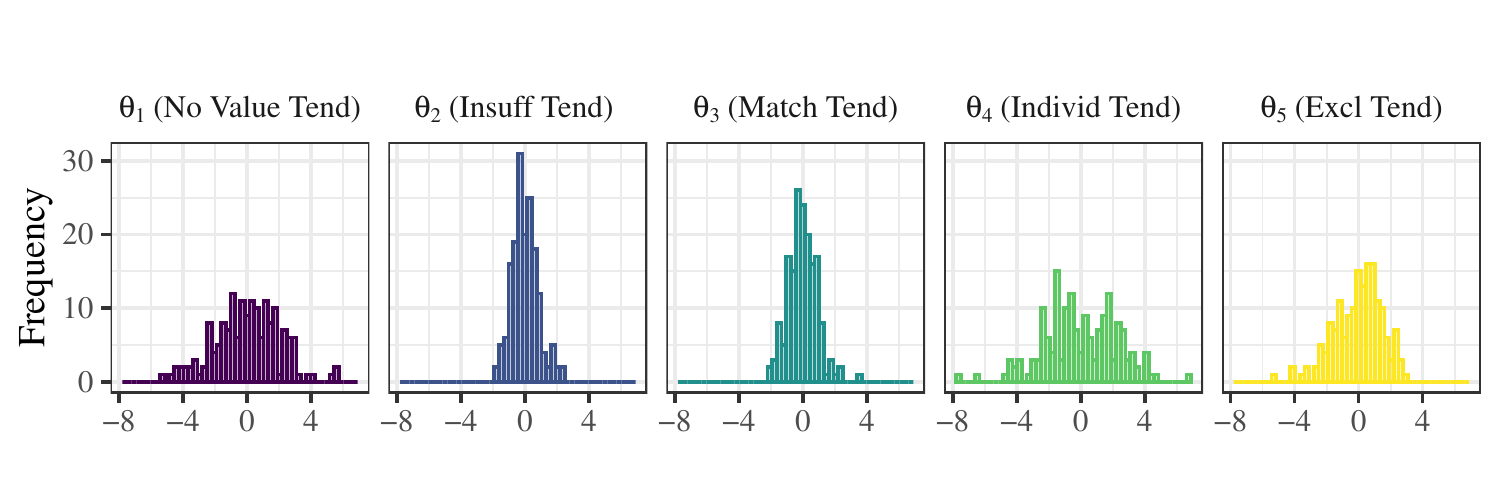}
		\caption{Distribution of $\theta$ point estimates under the binary decision process model. Greatest variation occurs in $\theta_1$, $\theta_4$, and $\theta_5$, corresponding to No Value, Individualization, and Exclusion tendencies, respectively.}
		\label{fig:hrthetas}
\end{figure}

\begin{figure}[ht]
		\centering
		\includegraphics[width=\linewidth]{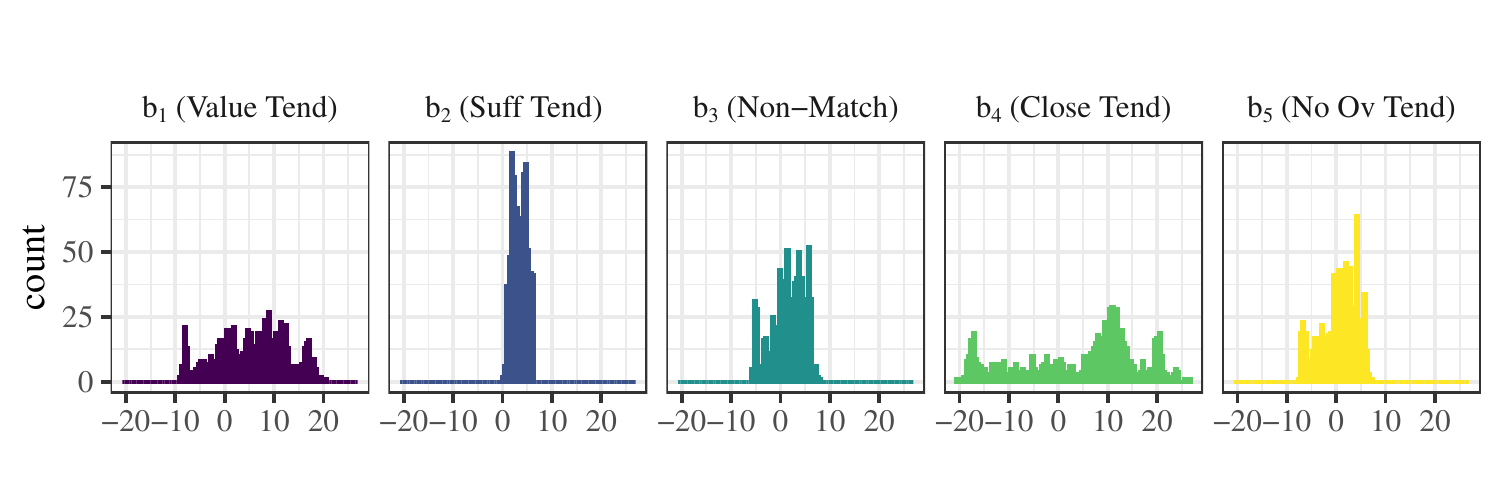}
		\caption{Distribution of $b$ point estimates under the binary decision process model. Greatest variation occurs in $b_1$, $b_4$, corresponding to Value and Close tendencies, respectively. Also note that $b$ values are more extreme than $\theta$ values.}
		\label{fig:hrbs}
\end{figure}

The posterior medians for each examiner and item were calculated, and the distribution of examiner parameters  (Figure~\ref{fig:hrthetas}) and item parameters (Figure~\ref{fig:hrbs}) are displayed as a whole. The item parameters are generally more extreme than the person parameters corresponding to the same decision (e.g. $\theta_1$ ranges from $\approx -6$ to $6$, while $b_1$ ranges from $\approx -10$ to $20$). This suggests that many of the responses are governed by item effects, rather than examiner tendencies. 

The greatest variation in person parameters occurs in $\theta_1$ (`no value' tendency), $\theta_4$ (conclusive tendency in matches) and $\theta_5$ (conclusive tendency in non-matches). Item parameters are most extreme in $b_1$ (tendency towards has value) and $b_4$ (inconclusive tendency in matches). For example, $b_{1,368}=-8.99$ and indeed all examiners agreed that item $368$ has no value; similarly $b_{4,166}=10.01$ and all examiners indeed agree that no individualization determination can be made for item $166$.  

Using probabilities calculated from the IRTree model estimates provides a way to assess the observed decisions in each examiner $\times$ item pair in light of other decisions that examiner made, and how other examiners evaluated that item. Inconclusives that are `expected' under the model can then be determined, along with which examiners often come to conclusions that are consistent with the model-based predictions. For example, an examiner whose responses often match the model-based predictions may be more proficient in recognizing when there is sufficient evidence to make a conclusive decision than an examiner whose responses do not match the model-based predictions. 

As one example, Examiner 55 decided Item 556  was a `Close' inconclusive, but Item 556 is a true non-match. Using posterior median estimates for $\theta_{k,55}$ and $b_{k,556}$ under the binary decision process model (where $k = 1, ..., 5$ and indexes each split in the tree), the probability of observing each response for this observation can be calculated: P(No Value) $< 0.005$, P(Individualization) $< 0.005$, P(Close) $= 0.20$, P(Insufficient) $< 0.005$, P(No Overlap) $= 0.01$ and P(Exclusion) $=0.78$. According to the model, the most likely outcome for this response is an exclusion. Since an inconclusive was observed instead, this response might be flagged as being due to examiner indecision. This process suggests a method for determining ``expected answers'' for each item using an IRTree approach, which we further discuss in Section \ref{ss:no-answer-key}. 
	
The estimated $\beta_{0k}$ and $\beta_{1k}$, with 90\% posterior intervals, are displayed in Table \ref{tab:regr-table}. Since the estimated $\beta_{1k}$'s all have posterior intervals that are entirely negative ($k=1, 2, 3$) or overlap zero ($k=4, 5$), we can infer that the identification tasks for true matches (e.g. $X_j = 1$ in Equation \ref{lltmequation}) tend to have lower $b_{jk}$ parameters than the true non-matches ($X_j=0$), leading to matching pairs being more likely fall along the left branches of the tree in Figure \ref{hyp-process-tree}. 

\begin{table}[]\small
\[
		\begin{array}{lrcrcrcrcrc}
			\toprule
			k:      & 1         &                & 2 &                      & 3      &                 & 4       &                & 5 \\
			\midrule
			\beta_{0k} & .87 & (.74, .99)      & 1.95 & (1.72, 2.19)      & .39 & (.13, .65)     & -.44 & (-.91,.024) & 4.58 & (3.60, 5.96)     \\
			\beta_{1k} & -.16 & (-.29, -.01) & -.27 & (-.46, -.09) & -.37 &  (-.55, -.2) & .19 & (-.15, .53)  & .06 & (-.35,.45) \\
			\bottomrule
		\end{array}
\]
\caption{Regression coefficients (with 90\% posterior intervals) for each of the five nodes in the IRTree model. }
	\label{tab:regr-table}
\end{table}
	
We also note that the IRTrees approach is compatible with the joint models for correctness and reported difficulty introduced in Section \ref{ss:reported-difficulty-model}. By replacing the Rasch model for correctness with an IRTree model, \citet{luby2019thesis} demonstrates that reported difficulty is related to IRTree branch propensities ($\theta_{ik} - b_{jk}$), with items tending to be rated as more difficult when the IRTree branch propensities are near zero.

Moreover, examiners are likely to use different thresholds for reporting difficulty, just as they do for coming to source evaluations \citep{aaasReport, ulery2017factors}; the IRTrees analysis above has been helpful in making these differing thresholds more explicit. In the same way, the IRTrees analysis of reported difficulty may lead to insights about how examiners decide how difficult an identification task is.

\subsection{Generating Answer Keys from Unscored Responses}
\label{ss:no-answer-key}

Generating evidence to construct test questions is both time-consuming and difficult. The methods introduced in this section provide a way to use evidence collected in non-controlled settings, for which ground truth is unknown, for testing purposes. Furthermore, examiners should receive feedback not only when they make false identifications or exclusions, but also if they make `no value' or `inconclusive' decisions when most examiners are able to come to a conclusive determination (or vice-versa). It is therefore important to distinguish when no value, inconclusive, individualization, and exclusion responses are expected in a forensic analysis. 

There are also existing methods for `IRT without an answer key', for example the cultural consensus theory (CCT) approach \citep{batchelder1988test, oravecz2014bayesian}. CCT was designed for situations in which a group of respondents shares some knowledge or beliefs in a domain area which is unknown to the researcher or administrator \citep[similar approaches have been applied to ratings of extended response test items, e.g.][]{casabianca2016hierarchical}. CCT then estimates the expected answers to the items provided to the group. We primarily focus on comparing the Latent Truth Rater Model (LTRM), a CCT model for ordinal categorical responses \citep{anders2015cultural}, to an IRTree-based approach. 

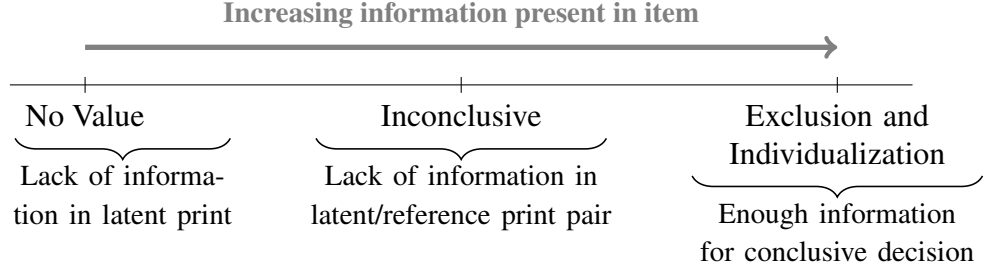
\begin{figure}[h]
	\centering
	\begin{tikzpicture}[Brace/.style args = {#1}{semithick, decorate, decoration={brace,#1,raise=6pt, amplitude = 8pt,
			pre=moveto,pre length=2pt,post=moveto,post length=2pt,}}]
	\usetikzlibrary{arrows}
	
	\draw (0,0) -- (12,0);
	\draw[->, color=gray, line width = 2pt] (1,.5) -> (11,.5);
	\foreach \x in {1,6,11}
	\draw (\x cm,3pt) -- (\x cm,-3pt);
	
	\draw (1,0) node[below=3pt] {No Value} node[above=3pt] {};
	\draw[Brace=mirror] (0,-.5) -- node[align = center, below=12pt, text width = 4cm, midway]{ \small  Lack of information in latent print} (3,-.5);
	\draw (3,0) node[below=3pt] {} node[above=3pt] {};
	\draw (6,0) node[below=3pt] {Inconclusive} node[above=3pt] {};
	\draw (6,.5) node[below=3pt] {} node[above=3pt] {\small \color{gray} \bf Increasing information present in item};
	\draw[Brace=mirror] (4,-.5) -- node[align = center, below=12pt, text width = 4cm, midway]{ \small Lack of information in latent/reference print pair} (8,-.5);
	\draw (9,0) node[below=3pt] {} node[above=3pt] {};
	\draw (11,0) node[below=3pt, align = center, text width = 3cm] {Exclusion and Individualization} node[above=3pt] {};
	\draw[Brace=mirror] (9,-1) -- node[align = center, below=12pt, text width = 4cm, midway]{\small Enough information for conclusive decision} (13,-1);
	\end{tikzpicture}
	\caption{FBI Black Box responses on a `conclusiveness' scale. } 
	\label{nv-conc-scale}
\end{figure}

Although the individualization/exclusion scale in Figure~\ref{bipolar-scale} could be used to generate an answer key for the source evaluations (i.e. individualization, exclusion, or inconclusive), it would not be possible to determine an answer key for the latent print value assessments (i.e. no value vs has value). Instead, a `conclusiveness' scale, Figure~\ref{nv-conc-scale}, can be used. This scale does not distinguish between same source and different source prints, but does allow for the inclusion of no value responses on the scale.  Using an answer key from this scale, alongside the same-source/different-source information provided by the FBI, provides a complete picture of what the expected answers are: An answer key generated for items placed on the scale of Figure~\ref{nv-conc-scale} identifies which items are expected to generate conclusive, vs. inconclusive answers; for the conclusive items, same-source pairs should be individualizations and  different-source pairs should be exclusions. 

\subsubsection{Models}

We fit four models to the Black Box Data: (1) The LTRM \citep{anders2015cultural}, (2) an adapted LTRM using a cumulative logits model (C-LTRM), (3) an adapted LTRM based using an adjacent logits model (A-LTRM), and (4) an IRTree model. Each of the four models is detailed below.

\subsubsection*{Latent Truth Rater Model}

Let $Y_{ij}=c$ denote examiner $i$'s categorical response to item $j$, where $c=1$ is the response ``No Value'', $c=2$ is the response ``Inconclusive'', and $c=3$ is the response ``Conclusive''. Key features of the LTRM in our context are $T_j$, the latent ``answer key'' for item $j$, and $\gamma_c$ ($c=1,2$), the category boundaries between `No Value' vs. `Inconclusive', and for `Inconclusive' vs. `Conclusive', respectively. Each examiner draws a latent appraisal of each item ($Z_{ij}$), which is assumed to follow a normal distribution with mean $T_j$ (the `true' location of item $j$) and precision $\tau_{ij}$, which depends on both examiner competency ($E_i$) and item difficulty ($\lambda_j$) (that is, $\tau_{ij} = \frac{E_i}{\lambda_j}$). If every examiner uses the `true' category boundaries, then if $Z_{ij} \le \gamma_1$ then $Y_{ij} =$ `No Value', if $ \gamma_1 \le Z_{ij} \le \gamma_2$ then $Y_{ij} =$ `Inconclusive', and if $ Z_{ij} \ge \gamma_2$ then $Y_{ij} = $`Conclusive'. Individuals, however, might use a biased form of the category thresholds, and so individual category thresholds, $\delta_{i,c} = a_i \gamma_c + b_i$, are defined, where $a_i$ and $b_i$ are examiner scale and shift biasing parameters, respectively. That is, $a_i$ shrinks or expands the category thresholds for examiner $i$, and $b_i$ shifts the category thresholds to the left or right. The model is thus
\begin{align}
P(Y_{ij} = \text{No Value}) &= P(Z_{ij} \le \delta_{i, 1}) = P(T_j + \epsilon_{ij} \le a_i \gamma_1 + b_i) = F(a_i \gamma_1 + b_i)
\end{align}
\begin{align}
P(Y_{ij} = \text{Inconclusive}) = P(\delta_{i, 1} < Z_{ij} \le \delta_{i, 2}) &= P(a_i \gamma_1 + b_i \le T_j + \epsilon_{ij} \le a_i \gamma_2 + b_i) \\
&= F(a_i \gamma_2 + b_i) - F(a_i\gamma_1 + b_i)
\end{align}
\begin{align}
P(Y_{ij} = \text{Conclusive}) = P(Z_{ij} > \delta_{i, 2}) = P(T_j + \epsilon_{ij} > a_i \gamma_2 + b_i) = 1- F(a_i \gamma_2 + b_i) ,
\end{align}
where $F(u)$ is the CDF of a normal variable with mean $T_j$ and precision $\tau_{ij}$. The likelihood of the data under the LTRM is then
\begin{equation}
L(\boldsymbol{Y}|\boldsymbol{T,a,b,\gamma, E, \lambda}) = \prod_I \prod_J [F(\delta_{i, y_{ij}}) - F(\delta_{i, y_{ij} -1})] ,
\label{eq:ltrm-likelihood}
\end{equation}
where $\delta_{i, 0} = - \infty$, $\delta_{i,3} = \infty$, and $\delta_{i,c} = a_i \gamma_c + b_i$.  We next consider adaptations of the LTRM to a logistic modeling framework, with some simplifying assumptions on the LTRM parameters.

\subsubsection*{Adapted LTRM as a Cumulative Logits Model (C-LTRM)}

The original LTRM (Equation~\ref{eq:ltrm-likelihood}) is a cumulative-probits model, and is therefore more closely related to more standard IRT models than it might seem at first glance. Specifically, if (1) the latent appraisals ($Z_{ij}$) are modeled with a logistic instead of a normal distribution, (2) it is assumed that $\tau_{ij} =\frac{E_i}{\lambda_j} = 1$ for all $i, j$, and (3) it is assumed $a_i = 1$ for all $i$, then the model collapses into a more familiar cumulative logits IRT model,
\begin{equation}
\log \frac{P(Y_{ij} \le c)}{P(Y_{ij} > c)} = b_i - T_j + \gamma_c.
\end{equation}
This transformed model has the same form as the Graded Response Model \citep{samejima_estimation_nodate}. Relaxing the assumption that $a_i = 1$, a cumulative logits model with a scaling effect for each person on the item categories is obtained, which we call the cumulative-logits LTRM (C-LTRM),
\begin{equation}
\log \frac{P(Y_{ij} \le c)}{P(Y_{ij} > c)} = b_i - T_j + a_i \gamma_c.
\label{eq:c-ltrm-logit}
\end{equation}
The likelihood for the data under Equation~\ref{eq:c-ltrm-logit} is
\begin{equation}
L(\boldsymbol{Y} | \boldsymbol{a, b, T, \gamma}) = \prod_I \prod_J \left[ \frac{\exp(b_i -  T_j + a_i \gamma_c)}{1+ \exp(b_i -  T_j + a_i \gamma_c)} - \frac{\exp(b_i -  T_j + a_i \gamma_{c-1})}{1+ \exp(b_i -  T_j + a_i \gamma_{c-1})} \right] ,
\end{equation}
where $\gamma_0 = -\infty$ and $\gamma_C = \infty$.  

\subsubsection*{Adapted LTRM as an Adjacent Category Logits Model (A-LTRM)}

Making the same assumptions as above, $P(Y_{ij} = c)$ could instead be expressed using an adjacent-categories logit model,
\begin{equation}
\log \frac{P(Y_{ij} = c)}{P(Y_{ij} = c-1)} = b_i - T_j + \gamma_c ,
\end{equation}
which takes the same form as the Rating Scale Model \citep{andrich1978application}. The RSM has nice theoretical properties due to the separability of $T_j$ and $b_i$ in the likelihood, and re-casting the LTRM as an adjacent-categories model opens the possibility of more direct theoretical comparisons between models. Relaxing the assumption that $a_i = 1$, a generalized adjacent-categories logit model with a scaling effect for each person on the item categories is obtained, which we call the adjacent-logits LTRM (A-LTRM),
\begin{equation}
\log \frac{P(Y_{ij} = c)}{P(Y_{ij} = c-1)} = b_i - T_j + a_i \gamma_c.
\end{equation}
The likelihood is then
\begin{equation}
L(\boldsymbol{Y} | \boldsymbol{a, b, T, \gamma}) = \prod_I \prod_J  \frac{\exp(b_i -  T_j + a_i \gamma_c)}{1+ \exp(b_i -  T_j + a_i \gamma_c)}.
\end{equation}

\subsubsection*{IRTree for answer key generation}

For comparison, we also consider a simplified IRTree model for answer key generation, which does not include the reason provided for inconclusive responses (as the model in Section~\ref{ss:structured-responses} did). This simplification was made for two reasons: first, this simplified IRTree model allows us to make inferences on the `conclusiveness' scale in Figure~\ref{nv-conc-scale}, facilitating comparison with the CCT model; second, the reasons provided for inconclusive responses are relatively inconsistent. Indeed, in a follow-up study done by the FBI \citep{ulery2012repeatability}, 72 Black Box study participants were asked to re-assess 25 items. 85\% of no value assessments, 90\% of exclusion evaluations, 68\% of inconclusive responses, and 89\% of individualization evaluations were repeated; while only 44\% of `Close', 21\% of `Insufficient', and 51\% of `No Overlap'  responses were repeated. Inconclusive reasoning thus varies more \textit{within} examiners than the source evaluations, and a generated answer key containing reasons for inconclusives may not be reliable or consistent across time.

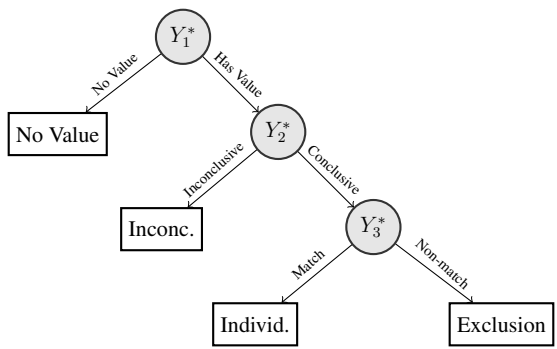
\begin{figure}[h]
	\centering
	\begin{tikzpicture}[scale = .7, transform shape]
	\tikzstyle{latent}=[circle, minimum size = 9mm, thick, draw =black!80, node distance = 15mm, fill = black!10]
	\tikzstyle{observed}=[rectangle, minimum size = 8mm, thick, draw = black!100, node distance = 15mm]
	\node[latent] (y1) { $Y^*_{1}$ };
	\node[observed] (nv) [below left = of y1] {No Value};
	\node[latent] (y2) [below right = of y1] { $Y^*_{2}$ };
	\node[observed] (inc) [below left = of y2] {Inconc.};
	\node[latent] (y3) [below right = of y2] { $Y^*_{3}$ };
	\node[observed] (ind) [below left = of y3] {Individ.};
	\node[observed] (ex) [below right = of y3] {Exclusion};
	\draw[->] (y1) -> (nv) node [midway, rotate=42, anchor = south]  {\scriptsize No Value};
	\draw[->] (y1) -> (y2) node [midway, rotate=-42, anchor = south]  {\scriptsize Has Value};
	\draw[->] (y2) -> (inc) node [midway, rotate=42, anchor = south]  {\scriptsize Inconclusive};
	\draw[->] (y2) -> (y3) node [midway, rotate=-42, anchor = south]  {\scriptsize Conclusive};
	\draw[->] (y3) -> (ind) node [midway, rotate=42, anchor = south]  {\scriptsize Match};
	\draw[->] (y3) -> (ex) node [midway, rotate=-42, anchor = south]  {\scriptsize Non-match};
	\end{tikzpicture}
	\caption{The \textit{answer key} IRtree}
	\label{answer-key-tree}
\end{figure}

The tree structure for the simplified IRTree model is shown in Figure~\ref{answer-key-tree}. The first internal node ($Y_{1}^*$) represents the value assessment, the second internal node ($Y_{2}^*$) represents the conclusive decision, and the third internal node represents the individualization/exclusion decision. Note that $Y_3^*$ is not a part of the conclusiveness scale in Figure~\ref{nv-conc-scale}, and thus provides additional information beyond the `conclusiveness' answer key. 

\subsubsection{Results}\label{ss:no-answer-key-results}

We focus on comparing the answer keys generated by each of the models.  As a simple baseline answer key, we also calculate the modal response for each item using the observed responses. Unlike the IRTree and LTRM approaches, this baseline answer key does not account for different tendencies of examiners who answered each item; nor does it account for items being answered by different numbers of examiners. 
The LTRM, A-LTRM, and C-LTRM all estimate the answer key, a combination of $T_j$'s and $\gamma_c$'s, directly. The answer for item $j$ is `No Value' if $T_j < \gamma_1$, `Inconclusive' if $\gamma_1 < T_j < \gamma_2$ and `Conclusive' if $ T_j > \gamma_2$. 
For the IRTree model, an answer key was calculated based on what one would expect an `unbiased examiner' to respond. The response of a hypothetical unbiased examiner (i.e. $\theta_{ki}= 0$ for all $k$) to each question was predicted, using the estimated item parameters in each split.

There are thus five answer keys: (1) Modal answer key, (2) LTRM answer key, (3) C-LTRM answer key, (4) A-LTRM answer key, and (5) IRTree answer key. Each of the answer keys has three possible answers: no value, inconclusive, or conclusive.  Table~\ref{tab:answer-key-comp} shows the number of items (out of 744) that the answer keys disagreed upon. The most similar answer keys were the A-LTRM and C-LTRM, which only disagreed on six items: three that disagreed between inconclusive/conclusive and three that disagreed between no value and inconclusive. The original LTRM model most closely matched the modal answer, with the A-LTRM model disagreeing with the modal answer most often. 

\begin{table}[h]
	\centering
     \begin{tabular}{@{}l|cccll@{}}
		\toprule
		& \multicolumn{1}{l}{Modal} & \multicolumn{1}{r}{LTRM} & \multicolumn{1}{l}{C-LTRM} & A-LTRM & IRTree \\ \midrule
		Modal & 0                             & -                        & -                          & -      & -      \\
		LTRM      & 12                            & 0                        & -                          & -      & -      \\
		C-LTRM    & 48                            & 39                       & 0                          & -      & -      \\
		A-LTRM    & 52                            & 43                       & 6                          & 0      & -      \\
		IRTree    & 32                            & 24                       & 28                         & 34     & 0      \\ \bottomrule
	\end{tabular}
	\caption{The number of items whose answers disagreed among the five approaches to finding an answer key. The C-LTRM and A-LTRM most closely matched each other, and the original LTRM answer key most closely matched the modal answer.}
	\label{tab:answer-key-comp}
	\end{table}

Recall that the three possible answers were (1) `no value', (2) `inconclusive', or (3) `conclusive'. There were 48 items for which at least one of the models disagreed with the others.  The vast majority of these disagreements were between `no value' and `inconclusive' or `inconclusive' and `conclusive'. Of the 48 items in which models disagreed, only five items were rated to be conclusive by some models and no value by others. All of these five items were predicted to be `no value' by the LTRM, `inconclusive' by the A-LTRM and C-LTRM, and `exclusion' by the IRTree. Table~\ref{tab:five-items} shows the number of observed responses in each category for these five items and illuminates two problems with the LTRM approaches. First, the original LTRM strictly follows the modal response, even when a substantial number of examiners came to a different conclusion. In Question 665, for example, eight examiners were able to make a correct exclusion, while the LTRM still chose `no value' as the correct response. Second, the A-LTRM and C-LTRM models may rely too much on the ordering of outcomes. Both adapted LTRM models predicted these items to be inconclusives, yet most examiners who saw the items rated it as either a `no value' or `exclusion'. 

\begin{table}[h]
	\centering
	\begin{tabular}{@{}l|cccl@{}}
		\toprule
		Item ID & No Value & Inconclusive & Exclusion  \\ \midrule
		427 & 13                             & 3                        & 13                                 \\
		438      & 12                            & 3                       & 7                      \\
		443    & 7                           & 1                       & 6                        \\
		665    & 9                            & 4                       & 8                         \\
		668    & 14                            & 1                       & 11                         \\ \bottomrule
	\end{tabular}
	\caption{The number of observed responses in each category for the five items with a disagreement between no value and conclusive.}
	\label{tab:five-items}
\end{table}

Using a model-based framework to generate expected answers provides more robust answer keys than relying on the observed responses alone. Both IRTrees and a CCT-based approach allow for the estimation of person and item effects alongside an answer key. Furthermore, although the two approaches are formulated quite differently, they lead to similar generated answer keys in the Black Box data. This similarity is due to the conditional sufficient statistics for item location parameters being closely related in the two models \citep[see][for further details]{luby2019thesis}. 

For this setting, we prefer using the IRTree framework to analyze responses because it does not require the responses to be ordered and because each decision may be modeled explicitly. In addition, model fit comparisons using the Widely Applicalble AIC index \citep[WAIC,][]{vehtari2017practical,watanabe2010asymptotic}, as well as in-sample prediction error, prefer the IRTree model for this data; see Table~\ref{tab:ugt-waic}.

	\begin{table}[ht]
		\centering
		\caption{WAIC and in-sample prediction error for each of the four models. In order to compare the IRTree to the LTRM models -- which only predict no value, inconclusive, or conclusive responses --  individualizations and exclusions (i.e. $Y_{3}^*$ in Figure \ref{answer-key-tree}) were grouped together.}
		\label{tab:ugt-waic}
		\begin{tabular}{lrrr}
			\toprule
			Model & WAIC & SE & In-Sample Prediction Error \\
			\midrule
			LTRM & 40416 & 748 & 0.19 \\
			C-LTRM & 13976 & 175 & 0.14 \\
			A-LTRM & 14053 & 178 & 0.15\\
			IRTree & 12484 & 166 & 0.12 \\
			\bottomrule
		\end{tabular}
	\end{table}
	
\section{Discussion and Future Work}\label{s:discussion}

In this survey of recent advances in the psychometric analysis of forensic decision-making process data, we have applied a wide variety of models, including the Rasch model, Item Response Trees, and Cultural Consensus Models, to identification tasks in the FBI Black Box study of error rates in fingerprint examination.  Careful analysis of forensic decision-making processes unearths a series of sequential responses that to date have often been ignored, while the final decision is simply scored as either correct or incorrect. Standard IRT models applied to scored data, such as the Rasch model of Section \ref{ss:rasch}, provide substantial improvements over current examiner error rate studies: examiner proficiencies can be justifiably compared even if the examiners did not do the same identification tasks, and the influence of the varying difficulty of identification tasks can be seen in examiner proficiency estimates.
Additional modeling techniques are needed to account for the co-varying responses present in the form of reported difficulty (Section \ref{ss:covarying-responses}), the sequential nature of examiner decision-making (Section \ref{ss:structured-responses}), and the lack of an answer key for scoring `no value' and `inconclusive' responses (Section \ref{ss:no-answer-key}). See \citet{luby2019thesis} for further developments of all methods presented here. 

In our analyses, we found a number of interesting results with important implications for subjective forensic science domains. 
Taken together, the results presented here demonstrate the rich possibilities in accurately modeling the complex decision-making in fingerprint identification tasks. 

For instance, results from  Section~\ref{ss:reported-difficulty-results} show that there are differences among fingerprint examiners in how they report the difficulty of identification tasks, and that this behavior is not directly related to examiners' estimated proficiency. Instead, examiners tended to over-rate task difficulty when the task was of middling difficulty, and under-rate the difficulty of tasks that were either extremely easy or extremely hard. A similar effect also holds for the intermediate decisions in an IRTree analysis \citep{luby2019thesis}. 

Furthermore, we have shown that there is substantial variability among examiners in their tendency to make no value and inconclusive decisions, even after accounting for the variation in items they were shown (Section~\ref{ss:sequential-responses-results}). The variation in these tendencies could lead to additional false identifications (in the case of ``no value" evidence being further analyzed), or to guilty perpetrators going free (in the case of ``valuable" evidence \textit{not} being further analyzed). To minimize the variation in examiner decisions, examiners should receive feedback not only when they make false identifications or exclusions, but also when they make mistaken ‘no value’ or ‘inconclusive’ decisions.  Finally, in Section \ref{ss:no-answer-key}, we show how to use the data to infer which 'no value' or 'inconclusive' responses are likely to be mistaken.

Our analyses were somewhat limited by available data;
the Black Box study was designed to measure examiner performance without ascertaining how those decisions were made.  Privacy \& confidentiality considerations on behalf of the persons providing fingerprints for the study make it impossible for the FBI to share the latent and reference prints for each identification task; if they were available we expect meaningful item covariates could be generated, perhaps through image analysis.   Similar considerations on behalf of examiners preclude the possibility of demographic or background variables (e.g. nature of training, number of years in service, etc.) linked to individual examiners; auxiliary information such as examiners' annotations of selected features, or their clarity and correspondence determinations, is also not available. Each of these, if available, might help elucidate individual differences in examiner behavior and proficiency. 

We anticipate future collaboration with experts in human decision making to improve the models and with fingerprint domain experts to determine the type and amount of data that would be needed to make precise and accurate assessments of examiner proficiency and task difficulty. Finally, we expect a future line of work will be to consider what would be needed to connect error rates, statistical measures of uncertainty, and examiner behavior collected from standardized/idealized testing situations such as those discussed in this paper, with task performance by examiners in authentic forensic investigations.


\bibliographystyle{apalike}
\bibliography{luby-mazumder-junker-2019} 

\begin{thebibliography}{}

\bibitem[AAAS, 2017]{aaasReport}
AAAS (2017).
\newblock {Forensic Science Assessments: A quality and Gap Analysis - Latent
  Fingerprint Examination}.
\newblock Technical report, (prepared by William Thompson, John Black, Anil
  Jain, and Joseph Kadane).

\bibitem[Anders and Batchelder, 2015]{anders2015cultural}
Anders, R. and Batchelder, W.~H. (2015).
\newblock Cultural consensus theory for the ordinal data case.
\newblock {\em Psychometrika}, 80(1):151--181.

\bibitem[Andrich, 1978]{andrich1978application}
Andrich, D. (1978).
\newblock Application of a psychometric rating model to ordered categories
  which are scored with successive integers.
\newblock {\em Applied psychological measurement}, 2(4):581--594.

\bibitem[Batchelder and Romney, 1988]{batchelder1988test}
Batchelder, W.~H. and Romney, A.~K. (1988).
\newblock Test theory without an answer key.
\newblock {\em Psychometrika}, 53(1):71--92.

\bibitem[Bécue et~al., 2019]{becue2019fingermarks}
Bécue, A., Eldridge, H., and Champod, C. (2019).
\newblock Fingermarks and other body impressions – a review (august 2016 –
  june 2019).

\bibitem[Casabianca et~al., 2016]{casabianca2016hierarchical}
Casabianca, J.~M., Junker, B.~W., and Patz, R.~J. (2016).
\newblock Hierarchical rater models.
\newblock In {\em Handbook of Item Response Theory, Volume One}, pages
  477--494. Chapman and Hall/CRC.

\bibitem[{De Boeck} and Partchev, 2012]{deboeck2012statsoft}
{De Boeck}, P. and Partchev, I. (2012).
\newblock Irtrees: Tree-based item response models of the glmm family.
\newblock {\em Journal of Statistical Software, Code Snippets}, 48(1):1--28.

\bibitem[de~Boeck and Wilson, 2004]{eirtbook}
de~Boeck, P. and Wilson, M. (2004).
\newblock {\em Explanatory Item Response Models: A generalized linear and
  nonlinear approach}.
\newblock Springer, New York.

\bibitem[Dror and Langenburg, 2019]{dror2019cannot}
Dror, I.~E. and Langenburg, G. (2019).
\newblock `cannot decide': The fine line between appropriate inconclusive
  determinations versus unjustifiably deciding not to decide.
\newblock {\em Journal of forensic sciences}, 64(1):10--15.

\bibitem[Evett and Williams, 1996]{evett1996review}
Evett, I. and Williams, R. (1996).
\newblock A review of the sixteen point fingerprint standard in england and
  wales.
\newblock {\em Journal of Forensic Identification}, 46:49--73.

\bibitem[Ferrando and Lorenzo-Seva, 2007]{ferrando_item_2007}
Ferrando, P.~J. and Lorenzo-Seva, U. (2007).
\newblock An {Item} {Response} {Theory} {Model} for {Incorporating} {Response}
  {Time} {Data} in {Binary} {Personality} {Items}.
\newblock {\em Applied Psychological Measurement}, 31(6):525--543.

\bibitem[Fischer, 1973]{fischer1973linear}
Fischer, G.~H. (1973).
\newblock The linear logistic test model as an instrument in educational
  research.
\newblock {\em Acta psychologica}, 37(6):359--374.

\bibitem[Fischer and Molenaar, 2012]{raschbook}
Fischer, G.~H. and Molenaar, I.~W. (2012).
\newblock {\em Rasch models: Foundations, recent developments, and
  applications}.
\newblock Springer Science \& Business Media, New York.

\bibitem[Gardner et~al., 2019]{gardner2019latent}
Gardner, B.~O., Kelley, S., and Pan, K.~D. (2019).
\newblock Latent print proficiency testing: An examination of test respondents,
  test-taking procedures, and test characteristics.
\newblock {\em Journal of forensic sciences}.

\bibitem[Garrett and Mitchell, 2017]{garrett2017proficiency}
Garrett, B.~L. and Mitchell, G. (2017).
\newblock The proficiency of experts.
\newblock {\em University of Pennsylvania Law Review}, 166:901.

\bibitem[Haber and Haber, 2014]{haber2014experimental}
Haber, R.~N. and Haber, L. (2014).
\newblock Experimental results of fingerprint comparison validity and
  reliability: A review and critical analysis.
\newblock {\em Science \& Justice}, 54(5):375--389.

\bibitem[Holland and Rosenbaum, 1986]{holland1986conditional}
Holland, P.~W. and Rosenbaum, P.~R. (1986).
\newblock Conditional association and unidimensionality in monotone latent
  variable models.
\newblock {\em The Annals of Statistics}, 14(4):1523--1543.

\bibitem[Janssen et~al., 2004]{janssen2004models}
Janssen, R., Schepers, J., and Peres, D. (2004).
\newblock Models with item and item group predictors.
\newblock In {\em Explanatory item response models}, pages 189--212. Springer.

\bibitem[Kellman et~al., 2014]{kellman2014}
Kellman, P.~J., Mnookin, J.~L., Erlikhman, G., Garrigan, P., Ghose, T.,
  Mettler, E., Charlton, D., and Dror, I.~E. (2014).
\newblock Forensic comparison and matching of fingerprints: using quantitative
  image measures for estimating error rates through understanding and
  predicting difficulty.
\newblock {\em PloS one}, 9(5):e94617.

\bibitem[Kerkhoff et~al., 2015]{kerkhoff2015}
Kerkhoff, W., Stoel, R., Berger, C., Mattijssen, E., Hermsen, R., Smits, N.,
  and Hardy, H. (2015).
\newblock Design and results of an exploratory double blind testing program in
  firearms examination.
\newblock {\em {Science \& Justice}}, 55(6):514 -- 519.

\bibitem[Langenberg, 2009]{langenberg2009performance}
Langenberg, G. (2009).
\newblock A performance study of the ace-v process: A pilot study to measure
  the accuracy, precision, reproducibility, repeatability, and biasability of
  conclusions resulting from the ace-v process.
\newblock {\em Journal of Forensic Identification}, 59(2):219.

\bibitem[Langenburg et~al., 2012]{langenburg2012informing}
Langenburg, G., Champod, C., and Genessay, T. (2012).
\newblock Informing the judgments of fingerprint analysts using quality metric
  and statistical assessment tools.
\newblock {\em Forensic science international}, 219(1-3):183--198.

\bibitem[Langenburg et~al., 2009]{langenburgchampodwertheim2009testing}
Langenburg, G., Champod, C., and Wertheim, P. (2009).
\newblock Testing for potential contextual bias effects during the verification
  stage of the ace-v methodology when conducting fingerprint comparisons.
\newblock {\em Journal of Forensic Sciences}, 54(3):571--582.

\bibitem[Lewandowski et~al., 2009]{lkjprior}
Lewandowski, D., Kurowicka, D., and Joe, H. (2009).
\newblock Generating random correlation matrices based on vines and extended
  onion method.
\newblock {\em Journal of multivariate analysis}, 100(9):1989--2001.

\bibitem[Liu et~al., 2015]{liu2015study}
Liu, S., Champod, C., Wu, J., Luo, Y., et~al. (2015).
\newblock Study on accuracy of judgments by chinese fingerprint examiners.
\newblock {\em Journal of Forensic Science and Medicine}, 1(1):33.

\bibitem[Luby, 2019a]{luby2019thesis}
Luby, A. (2019a).
\newblock {\em Accounting for Individual Differences among Decision-Makers with
  Applications in Forensic Evidence Evaluation}.
\newblock PhD thesis, Carnegie Mellon University. Available from:
  http://www.swarthmore.edu/NatSci/aluby1/files/luby-dissertation.pdf.

\bibitem[Luby, 2019b]{luby2019openforsci}
Luby, A. (2019b).
\newblock Decision-making in forensic identification tasks.
\newblock In Tyner, S. and Hofmann, H., editors, {\em Open Forensic Science in
  R}, chapter~8. rOpenSci Foundation, US.

\bibitem[Luby and Kadane, 2018]{luby2018proficiency}
Luby, A.~S. and Kadane, J.~B. (2018).
\newblock Proficiency testing of fingerprint examiners with bayesian item
  response theory.
\newblock {\em Law, Probability and Risk}, 17(2):111--121.

\bibitem[Max et~al., 2019]{max2019assessing}
Max, B., Cavise, J., and Gutierrez, R.~E. (2019).
\newblock Assessing latent print proficiency tests: Lofty aims, straightforward
  samples, and the implications of nonexpert performance.
\newblock {\em Journal of Forensic Identification}, 69(3):281--298.

\bibitem[Oravecz et~al., 2014]{oravecz2014bayesian}
Oravecz, Z., Vandekerckhove, J., and Batchelder, W.~H. (2014).
\newblock Bayesian cultural consensus theory.
\newblock {\em Field Methods}, 26(3):207--222.

\bibitem[Pacheco et~al., 2014]{pacheco2014miami}
Pacheco, I., Cerchiai, B., and Stoiloff, S. (2014).
\newblock Miami-dade research study for the reliability of the ace-v process:
  Accuracy \& precision in latent fingerprint examinations.
\newblock {\em Unpublished report.}, pages 2--5.

\bibitem[{President’s Council of Advisors on Science and Technology},
  2016]{pcast}
{President’s Council of Advisors on Science and Technology} (2016).
\newblock Forensic science in criminal courts: Ensuring scientific validity of
  feature-comparison methods.
\newblock Technical report, Executive Office of The President’s Council of
  Advisors on Science and Technology, Washington DC.

\bibitem[{R Core Team}, 2013]{Rman}
{R Core Team} (2013).
\newblock {\em R: A Language and Environment for Statistical Computing}.
\newblock R Foundation for Statistical Computing, Vienna, Austria.

\bibitem[Rasch, 1960]{rasch1960studies}
Rasch, G. (1960).
\newblock {\em Probabilistic models for some intelligence and attainment
  tests}.
\newblock University of Chicago Press, Chicago.

\bibitem[Saks and Koehler, 2008]{saks2008individualization}
Saks, M.~J. and Koehler, J.~J. (2008).
\newblock The individualization fallacy in forensic science evidence.
\newblock {\em Vand. L. Rev.}, 61:199.

\bibitem[Samejima, 1969]{samejima_estimation_nodate}
Samejima, F. (1969).
\newblock {Estimation} {of} {Latent} {Ability} {Using} {a} {Response} {Pattern}
  {of} {Graded} {Scores}.
\newblock page~97.

\bibitem[{Stan Development Team}, 2018a]{rstan}
{Stan Development Team} (2018a).
\newblock {\em {RStan}: the {R} interface to {Stan}}.
\newblock R package version 2.18.2.

\bibitem[{Stan Development Team}, 2018b]{stan}
{Stan Development Team} (2018b).
\newblock {\em Stan Modeling Language Users Guide and Reference Manual}.

\bibitem[Tangen et~al., 2011]{tangen2011identifying}
Tangen, J.~M., Thompson, M.~B., and McCarthy, D.~J. (2011).
\newblock Identifying fingerprint expertise.
\newblock {\em Psychological science}, 22(8):995--997.

\bibitem[Taylor et~al., 2012]{nist2012latent}
Taylor, M.~K., Kaye, D.~H., Busey, T., Gische, M., LaPorte, G., Aitken, C.,
  Ballou, S.~M., Butt, L., Champod, C., Charlton, D., et~al. (2012).
\newblock Latent print examination and human factors: Improving the practice
  through a systems approach. report of the expert working group on human
  factors in latent print analysis.
\newblock Technical report, U.S. Department of Commerce, National Institute of
  Standards and Technology (NIST).

\bibitem[Thissen, 1983]{thissen_9_1983}
Thissen, D. (1983).
\newblock {Timed} {Testing}: {An} {Approach} {Using} {Item} {Response}
  {Theory}.
\newblock In Weiss, D.~J., editor, {\em New {Horizons} in {Testing}},
  chapter~9, pages 179--203. Academic Press, San Diego.

\bibitem[Ulery et~al., 2011]{ulery2011accuracy}
Ulery, B.~T., Hicklin, R.~A., Buscaglia, J., and Roberts, M.~A. (2011).
\newblock Accuracy and reliability of forensic latent fingerprint decisions.
\newblock {\em Proceedings of the National Academy of Sciences},
  108(19):7733--7738.

\bibitem[Ulery et~al., 2012]{ulery2012repeatability}
Ulery, B.~T., Hicklin, R.~A., Buscaglia, J., and Roberts, M.~A. (2012).
\newblock Repeatability and reproducibility of decisions by latent fingerprint
  examiners.
\newblock {\em PloS one}, 7(3):e32800.

\bibitem[Ulery et~al., 2014]{ulery2014measuring}
Ulery, B.~T., Hicklin, R.~A., Roberts, M.~A., and Buscaglia, J. (2014).
\newblock Measuring what latent fingerprint examiners consider sufficient
  information for individualization determinations.
\newblock {\em PloS one}, 9(11):e110179.

\bibitem[Ulery et~al., 2017]{ulery2017factors}
Ulery, B.~T., Hicklin, R.~A., Roberts, M.~A., and Buscaglia, J. (2017).
\newblock Factors associated with latent fingerprint exclusion determinations.
\newblock {\em Forensic science international}, 275:65--75.

\bibitem[van~der Linden, 2006]{van_der_linden_lognormal_2006}
van~der Linden, W.~J. (2006).
\newblock A {Lognormal} {Model} for {Response} {Times} on {Test} {Items}.
\newblock {\em Journal of Educational and Behavioral Statistics},
  31(2):181--204.

\bibitem[van~der Linden et~al., 2010]{van2010irt}
van~der Linden, W.~J., Klein~Entink, R.~H., and Fox, J.-P. (2010).
\newblock Irt parameter estimation with response times as collateral
  information.
\newblock {\em Applied Psychological Measurement}, 34(5):327--347.

\bibitem[Vehtari et~al., 2017]{vehtari2017practical}
Vehtari, A., Gelman, A., and Gabry, J. (2017).
\newblock Practical bayesian model evaluation using leave-one-out
  cross-validation and waic.
\newblock {\em Statistics and Computing}, 27(5):1413--1432.

\bibitem[Watanabe, 2010]{watanabe2010asymptotic}
Watanabe, S. (2010).
\newblock Asymptotic equivalence of bayes cross validation and widely
  applicable information criterion in singular learning theory.
\newblock {\em Journal of Machine Learning Research}, 11(Dec):3571--3594.

\bibitem[Wertheim et~al., 2006]{wertheim2006report}
Wertheim, K., Langenburg, G., and Moenssens, A. (2006).
\newblock A report of latent print examiner accuracy during comparison training
  exercises.
\newblock {\em Journal of forensic identification}, 56(1):55.

\end{thebibliography}

\end{document}